\documentclass{aa}

\usepackage{natbib}
\bibpunct{(}{)}{;}{a}{}{,} % to follow the A&A style
\usepackage{graphics}   % standard graphics specifications
\usepackage{graphicx}   % alternative graphics specificati.jpgons
\usepackage{amsmath}
\usepackage{epstopdf}
\usepackage[varg]{txfonts}
\usepackage{pdflscape}
\usepackage{hyperref}   %link
\usepackage[flushleft]{threeparttable}

\hypersetup{
    bookmarks=true,         % show bookmarks bar?
    unicode=false,          % non-Latin characters in Acrobat?s bookmarks
    colorlinks=true,       % false: boxed links; true: colored links
    linkcolor=blue,          % color of internal links (change box color with linkbordercolor)
    citecolor=blue,        % color of links to bibliography
    filecolor=blue,      % color of file links
    urlcolor=blue           % color of external links
}

\begin{document} 

   \title{Pre-supernova mixing in CEMP-no source stars}

   \author{
   Arthur Choplin
          \inst{1},
          Sylvia Ekstr\"{o}m
          \inst{1},
           Georges Meynet
           \inst{1},
          Andr\'{e} Maeder
          \inst{1},
          Cyril Georgy
          \inst{1},
          \and
          Raphael Hirschi
          \inst{2,3,4}
                    }

 \authorrunning{Choplin et al.}

        \institute{Geneva Observatory, University of Geneva, Maillettes 51, CH-1290 Sauverny, Switzerland\\
        e-mail: arthur.choplin@unige.ch
        \and Astrophysics Group, Lennard-Jones Labs 2.09, Keele University, ST5 5BG, Staffordshire, UK
        \and Kavli Institute for the Physics and Mathematics of the Universe (WPI), University of Tokyo, 5-1-5
Kashiwanoha, Kashiwa, 277-8583, Japan
        \and UK Network for Bridging the Disciplines of Galactic Chemical Evolution (BRIDGCE)
                }
   \date{Received / Accepted}
 
  \abstract
  %Context
   {CEMP-no stars are long-lived low-mass stars with a very low iron content, overabundances of carbon and no or minor signs for the presence of s- or r-elements. Although their origin is still a matter of debate, they are often considered as being made of a material ejected by a previous stellar generation (source stars).}
   %Aims
   {We place constraints on the source stars from the observed abundance data of CEMP-no stars.}
   %Methods
   {We computed source star models of $20$, $32$, and $60\,M_\odot$ at $Z=10^{-5}$ with and without fast rotation. For each model we also computed a case with a late mixing event occurring between the hydrogen and helium-burning shell $\sim 200$ yr before the end of the evolution. This creates a partially CNO-processed zone in the source star. We use the $^{12}$C/$^{13}$C and C/N ratios observed on CEMP-no stars to put constraints on the possible source stars (mass, late mixing or not). Then, we inspect more closely the abundance data of six CEMP-no stars and select their preferred source star(s).}
   %Results
   {Four out of the six CEMP-no stars studied cannot be explained without the late mixing process in the source star. Two of them show nucleosynthetic signatures of a progressive mixing (due e.g. to rotation) in the source star. We also show that a $20\,M_\odot$ source star is preferred compared to one of $60\,M_\odot$ and that likely only the outer layers of the source stars were expelled to reproduce the observed $^{12}$C/$^{13}$C.}
  %Conclusions
   {The results suggest that (1) a late mixing process could operate in some source stars, (2) a progressive mixing, possibly achieved by fast rotation, is at work in several source stars, (3) $\sim 20\,M_\odot$ source stars are preferred compared to $\sim 60\,M_\odot$ ones, and (4) the source star might have preferentially experienced a low energetic supernova with large fallback.
   }

   \maketitle

        \titlerunning{Spinstars and CEMP-no stars}
        \authorrunning{A. Choplin et al.}

\section{Introduction}\label{intro}

Carbon-enhanced metal-poor (CEMP) stars belong to the class of iron-deficient stars, and present an excess of carbon compared to the classical metal-poor stars \citep[we
refer to][for a recent review of metal-poor stars]{frebel15}. The CEMP frequency rises as [Fe/H] decreases, with increasing distance from the Galactic plane and when moving from the inner to outer halo \citep{frebel06,carollo12,lee13}. The two criteria defining a CEMP star are\footnote{[X/Y] $=\log_{10}(N_\text{X} / N_\text{Y}) - \log_{10}(N_{\text{X}\odot} / N_{\text{Y}\odot})$ with $N_\text{X,Y}$ the number density of elements X and Y.} [Fe/H] $< -1.0$ and [C/Fe] $> 0.7$ \citep{aoki07}. Based on the amounts in s- and r-elements, a division of the CEMP class in four categories was made by \cite{beers05}.  CEMP-s stars have their surface enriched in s-elements synthesised thanks to the slow neutron-capture process. The main scenario explaining the peculiar abundances observed at the surface of the CEMP-s stars is the binary mass transfer scenario \citep{bisterzo10,lugaro12,abate15}, supported by radial velocity detection of a companion for most of these objects \citep{lucatello05,starkenburg14}. However, the binary frequency of the CEMP-s stars does not seem to reach 100 \% \citep{hansen16} so another process could be responsible for the formation of some CEMP-s stars. The second and third classes are the CEMP-r/s and CEMP-r \citep[e.g.][]{mcwilliam95, sneden03, goswami06, roederer14b}. Due to the correlation between the observed abundances in -s and -r/s stars, \cite{allen12} argued that the binary mass transfer scenario is also valid for CEMP-r/s stars, the r-elements being explained by a pre-existing source that polluted the molecular cloud, such as one or several Type II supernovae. The origin of the r-element-enrichment is however still largely debated. The fourth category, so-called CEMP-no ("no" for the absence of s- or r- elements), is of particular interest since it dominates at [Fe/H] $\lesssim -3$ \citep{aoki10,norris13}, allowing us to approach the primordial universe even closer. Their formation process likely differs from the one of the CEMP-s stars. Indeed, \cite{starkenburg14}, using Monte Carlo simulations to constrain the binary fraction and binary period, concluded that the complete CEMP-no data set is inconsistent with the binary properties of the CEMP-s class. The formation scenarios for the CEMP-no stars generally assume that these stars formed from a cloud that was enriched by a previous generation of stars hereafter referred to as source stars. An assumption often made is that one CEMP-no star comes from one source star. If the CEMP-no star has not experienced mixing from its birth to now, the observed abundances at the surface are the same as the ones in the cloud that formed the star. There are three broad categories of models proposed to explain the CEMP-no stars.

In the "mixing and fallback" scenario \citep{umeda02,umeda05,tominaga14} the sources of the peculiar abundances shown by the CEMP-no stars are faint supernovae from Population III (Pop. III) stars. The supernova is faint because part of the envelope falls back on the remnant black-hole. Some mixing in internal regions of the source star is assumed, allowing part of the inner chemical species, such as iron, to be nevertheless ejected in small quantities. The mass cut\footnote{The mass cut delimits the part of the star that is expelled from the part that is locked into the remnant.} and the mixed region are free parameters, adjusted for each CEMP-no star to reproduce the observed abundance pattern.

The "spinstar" scenario \citep{meynet06,meynet10,hirschi07,chiappini13,maeder14} states that the material constituting a CEMP-no star comes from a massive source star experiencing mass-loss and strong internal mixing, owing to an average-to-high rotation rate. 
In this scenario, the light elements (C to Si) come directly from the source star. The small amounts of heavier elements (e.g. Ca, Ti, Ni) have either been produced by a generation of stars preceding the source stars, or by the source star. In the latter case, a very small amount of heavy elements should be ejected. This can be achieved through the models of fall back and mixing invoked by \cite{umeda02}. In those models, small amounts of heavy elements made their path through the ejected material thanks to a mixing process assumed to occur at the time of the supernova explosion.
Interestingly, the spinstar scenario, using the yields of such fast-rotating stars in the context of a chemical evolution model for the halo can reproduce many observed characteristics of the chemical composition of normal halo stars \citep{chiappini13}. Thus, CEMP-no and normal halo stars might be due to the same type of stars but from different reservoirs. In the case of the CEMP-no star, the reservoir of matter from which the star forms is a pocket of matter enriched by the ejecta of one or perhaps two fast-rotating massive stars. The normal halo stars, on the other hand, would be formed from a well mixed reservoir enriched by many more stars of different generations. 

The ``two supernovae model'' \citep{limongi03} assumes that the peculiar composition of one of the CEMP-no stars, HE~0107-5240, can be explained by the concurrent pollution of two supernova events; for instance a supernova of quite low mass (about $15\,M_\odot$) that underwent a normal explosion and  a supernova of a massive enough star (about $35\,M_\odot$) that experienced a strong fallback that locked all the carbon-oxygen core in a compact remnant.

Presently, there is no strong argument favouring one scenario over another. 
In this paper, we propose to further investigate the spinstar scenario. We show that
if the mass-loss rates that are used in our models accurately describe reality, the chemical composition of some CEMP-no stars needs to be explained by some mixing occurring very late in the course of the evolution of the source stars, typically a few hundred years before the core collapse. This mixing would take place at the interface between the hydrogen and helium-burning shells, and is considered here in a parametric form and thus cannot be attributed to a particular physical phenomenon (convection, rotation, etc.). However, we discuss evidence for its presence on the basis of nuclear processes.
An additional transport process has already been invoked in \cite{eggenberger16} for instance, in order to reproduce the low degree of radial differential rotation of the red giants, revealed by asteroseismic measurements. 
We study here the distinct nucleosynthetic signature of four categories of models: (1) no rotation, no late mixing, (2) no rotation, late mixing, (3) rotation, no late mixing, and (4) rotation, late mixing. Then, we try to see whether or not the nucleosynthetic signatures of these models are found at the surface of the CEMP-no stars.

The physical ingredients are presented in Sect.~\ref{sec:2}. Sect.~\ref{sec:3} discusses the "[C/N] $- ^{12}\text{C}/^{13}\text{C}$ puzzle" that presents itself when confronting source-star models with observed CEMP-no stars. Sect.~\ref{sec:4} focuses on the late mixing process as a possible solution to the problem. A parametric study of this mixing is done in Sect.~\ref{sec:5}. Sect.~\ref{sec:6} highlights nucleosynthetic signatures of the different source-star models. CEMP-no stars are inspected individually in Sect.~\ref{sec:7}. A discussion and the conclusions are given in Sect.~\ref{sec:8} and \ref{sec:9}, respectively.

\section{Ingredients of the models}
\label{sec:2}

\begin{table*}
\scriptsize{
\caption{Properties of source-star models: model name (column 1) initial mass (column 2), $V_\text{ini}/V_\text{crit}$ (column 3), $\Omega_\text{ini}/\Omega_\text{crit}$ (column 4), initial equatorial velocity (column 5), initial composition (column 6), total lifetime (column 7), lifetime of the MS, helium, carbon, neon, oxygen, and silicon-burning phases (column $7 - 13$), mass of the model at the end of MS, helium-burning phase, and at the end of evolution (column $14 - 16$). \label{table:1}}
\begin{center}
\resizebox{18.0cm}{!} {
\begin{tabular}{l|c|ccc|c|ccccccc|ccc} 
\hline % inserts double horizontal lines
\hline % inserts single horizontal line
Model & $M_\text{ini}$ &  $V_\text{ini}/V_\text{crit}$  & $\Omega_\text{ini}/\Omega_\text{crit}$ & $V_\text{eq,ini}$  & Initial& $\tau_\text{life}$ &  $\tau_\text{MS}$& $\tau_\text{He}$  & $\tau_\text{C}$ & $\tau_\text{Ne}$ & $\tau_\text{O}$ & $\tau_\text{Si}$ & $M_\text{MS}$ & $M_\text{He}$ & $M_\text{final}$  \\ % table heading
 & [$M_\odot$]&   & & [km/s] &  composition & [Myr] & [Myr]  &[Myr] &[yr] &[day] &[day] &[day]  &[$M_\odot$]& [$M_\odot$] & [$M_\odot$]  \\
\hline % inserts single horizontal line
\multicolumn{16}{c}{No rotation}\\
\hline  
20s0    &  20           &               0               &       0               &  0 &    $\alpha$-enhanced       &       8.93    &       8.02    &       0.79    &       978                     & 168 & 318 & 3.1 &       19.99   &  19.98                &       19.98                    \\
32s0    & 32            &               0               &       0               &  0 &    $\alpha$-enhanced       &       5.78    &       5.24    &       0.48    &       124                     & 22 & 47 & 0.6 & 31.97   &  31.94                &       31.94                   \\
60s0    & 60            &               0               &       0               &  0 &    $\alpha$-enhanced       &       3.81    &       3.44    &       0.33    &       15                      & 7 & 7 & 0.5 &   59.87   &  59.81                &       59.80                    \\
 \hline
 \multicolumn{16}{c}{Rotation}\\
\hline
20s7    & 20            &               0.7             &       0.88            &  610 &  $\alpha$-enhanced       &       11.0    &       10.1    &       0.76    &       400                     & 277 & 128 & 1.4 &       19.84   &  19.84                &       19.50                   \\
32s7    & 32            &               0.7             &       0.88            &  680 &  $\alpha$-enhanced       &       7.14    &       6.61    &       0.47    &       45                      & 7 & 15 & 0.6 &  31.48   &  31.42                &       30.71           \\
60s7    & 60            &               0.7             &       0.88            &  770 &  $\alpha$-enhanced       &       4.69    &       4.33    &       0.32    &       5                       & 1 & 3 & 0.2 &   58.47   &  47.97& 47.65                 \\
\hline
\end{tabular}
}
\end{center}
}
\end{table*}

We first explore six rotating and non-rotating source-star models of $20$, $32$, and $60\,M_\odot$.  The metallicity is set to $Z = 10^{-5}$ ([Fe/H] $= -3.8$) and the initial rotation rate, $V/V_\text{crit}$\footnote{$V_\text{crit}$ is the velocity at the equator at which the gravitational acceleration is exactly compensated by the centrifugal force \citep[see][]{maeder00a}.} , is either $0$ or $0.7$.

The initial composition of metals (elements heavier than helium\footnote{The initial helium mass fraction $Y$ is calculated according to the relation $Y=Y_\text{p} + \Delta Y / \Delta Z \times Z$ where $Z$ is the metallicity, $Y_\text{p}$ the primordial helium abundance and $\Delta Y / \Delta Z = (Y_{\odot} - Y_\text{p})/Z_{\odot}$ the average slope of the helium-to-metal enrichment law.  We set $Y_\text{p}=0.248$, according to \cite{cyburt03}. We use $Z_{\odot}=0.014$ and $Y_{\odot} = 0.266$ as in \cite{ekstrom12}, derived from \cite{asplund05}. The initial helium mass fraction calculated, the initial mass fraction of hydrogen is then deduced from $1-Y-Z = 0.752$.}) is $\alpha$-enhanced. In this case, $^{12}$C, $^{16}$O, $^{20}$Ne and $^{24}$Mg are enhanced relative to iron \citep[for more details, we refer to Sect. \textsection 2.1 of][]{frischknecht16}. The initial mixture at such a very low metallicity is poorly known. We take here an $\alpha$-enhanced mixture for all the models, as taken for the low metallicity models of \cite{meynet06}, \cite{hirschi07} or \cite{frischknecht16} for example. Other initial mixtures cannot be excluded: The chemical heterogeneity of the interstellar medium (ISM) at very low metallicity may lead to different metal mixtures for the source stars. However, for most of the elements considered in this work, the abundances in the ejecta of the source-star models are so different from the initial ones that they depend very weakly on the initial composition.

The opacity tables were computed with the OPAL tool\footnote{\url{http://opalopacity.llnl.gov}}. They are complemented at low temperatures by the opacities from \cite{ferguson05}. The mass-loss rates are from \cite{kudritzki00} when $\log T_\text{eff} \geq 3.95$ and from \cite{jager88} when $\log T_\text{eff} < 3.95$.

   \begin{figure}
   \centering
       \includegraphics[scale=0.49]{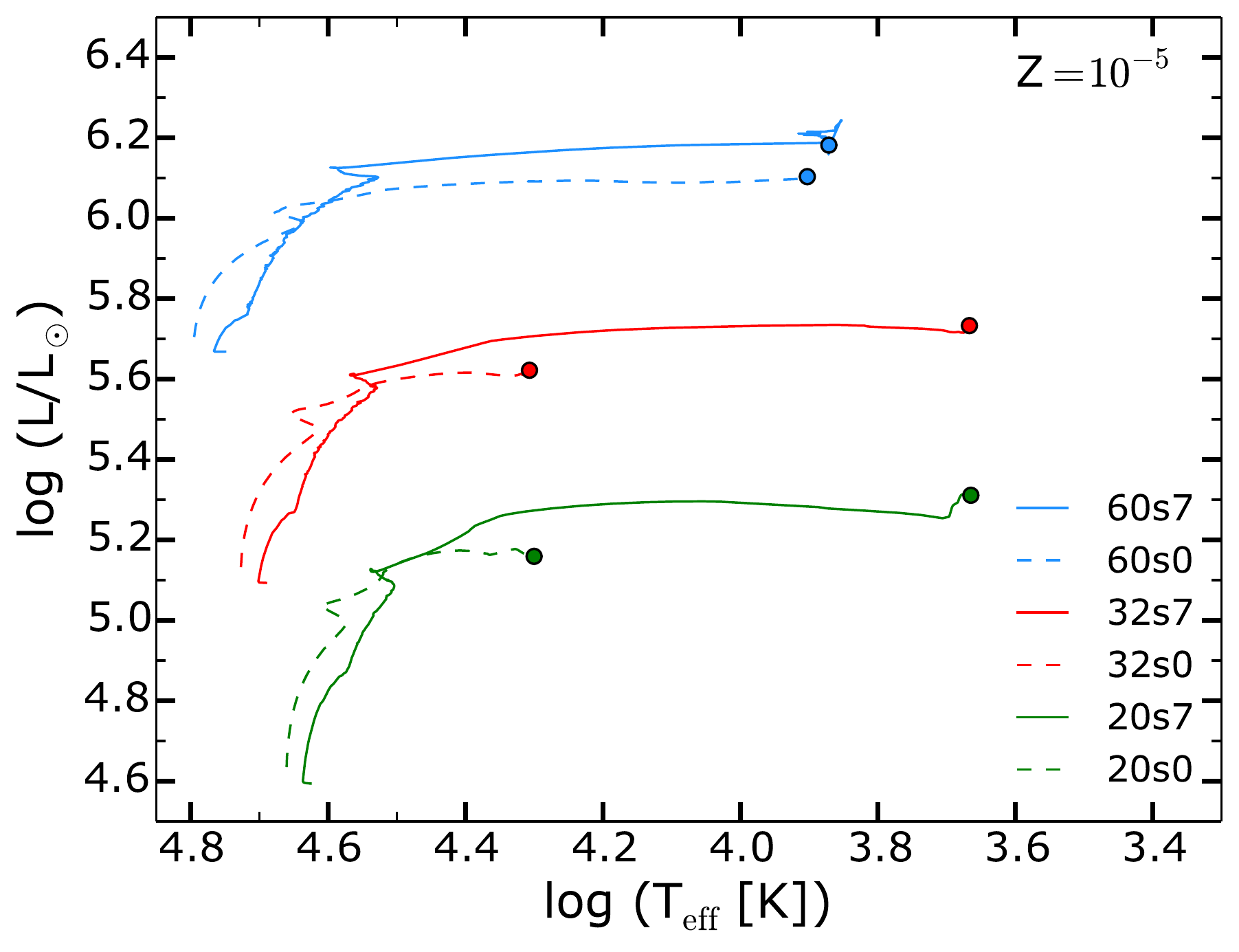}
   \caption{Tracks of the models of Table \ref{table:1} in the Hertzsprung-Russell diagram. 
   }
\label{HRD}
    \end{figure}

Among the physical ingredients needed to describe a star in differential rotation, the $D_\text{shear}$ coefficient is of major importance. This diffusion coefficient intervenes in the diffusion equation for the transport of chemical elements  in the differentially rotating layers. The $D_\text{shear}$ coefficient used in the present models is from \cite{talon97}. It is expressed as:
\begin{equation}
D_\text{shear} = f_\text{energ} \frac{H_\text{p}}{g \delta} \frac{K + D_\text{h}}{\left[\frac{\varphi}{\delta} \nabla_\mu \left(1+\frac{K}{D_\text{h}}\right) + \nabla_\text{ad} - \nabla_\text{rad}\right]}  \left( \frac{9\pi}{32} \Omega \frac{\text{d} \ln \Omega}{\text{d} \ln r} \right)^2
\label{dshearTZ}
,\end{equation}
where $K=\frac{4ac}{3\kappa}\frac{T^4 \nabla_\text{ad}}{\rho P \delta}$ is the thermal diffusivity, $D_\text{h}$ the diffusion coefficient for horizontal turbulence taken from \cite{zahn92}, and $f_\text{energ}$ the fraction of the excess energy in the shear that contributes to mixing (taken equal to 1).

We took the nuclear rates used in the Geneva grids \citep[see e.g.][]{ekstrom12, georgy13}. For the CNO cycle, they are mainly from \cite{angulo99}. Almost all the rates related to the Ne-Na Mg-Al chains are from \cite{hale02}. Only $^{20}$Ne($p,\gamma$)$^{21}$Na and $^{21}$Ne($p,\gamma$)$^{22}$Na are taken from \cite{angulo99} and \cite{iliadis01}, respectively. We have also taken into account our previous study where it is shown that the aluminum range observed in CEMP-no stars is either rather well reproduced or overestimated, depending on the nuclear rates used for the reactions involving $^{27}$Al \citep{choplin16pap}. Following this study, we took the rate of \cite{angulo99} for $^{26}$Mg($p,\gamma$)$^{27}$Al and the rates of \cite{cyburt10} for $^{27}$Al($p,\gamma$)$^{28}$Si and $^{27}$Al($p,\alpha$)$^{24}$Mg. Those rates favour a low synthesis of $^{27}$Al in the hydrogen-burning shell, that is likely needed to reproduce the observed aluminium distribution. 

\begin{figure}
   \centering
      \includegraphics[scale=0.25]{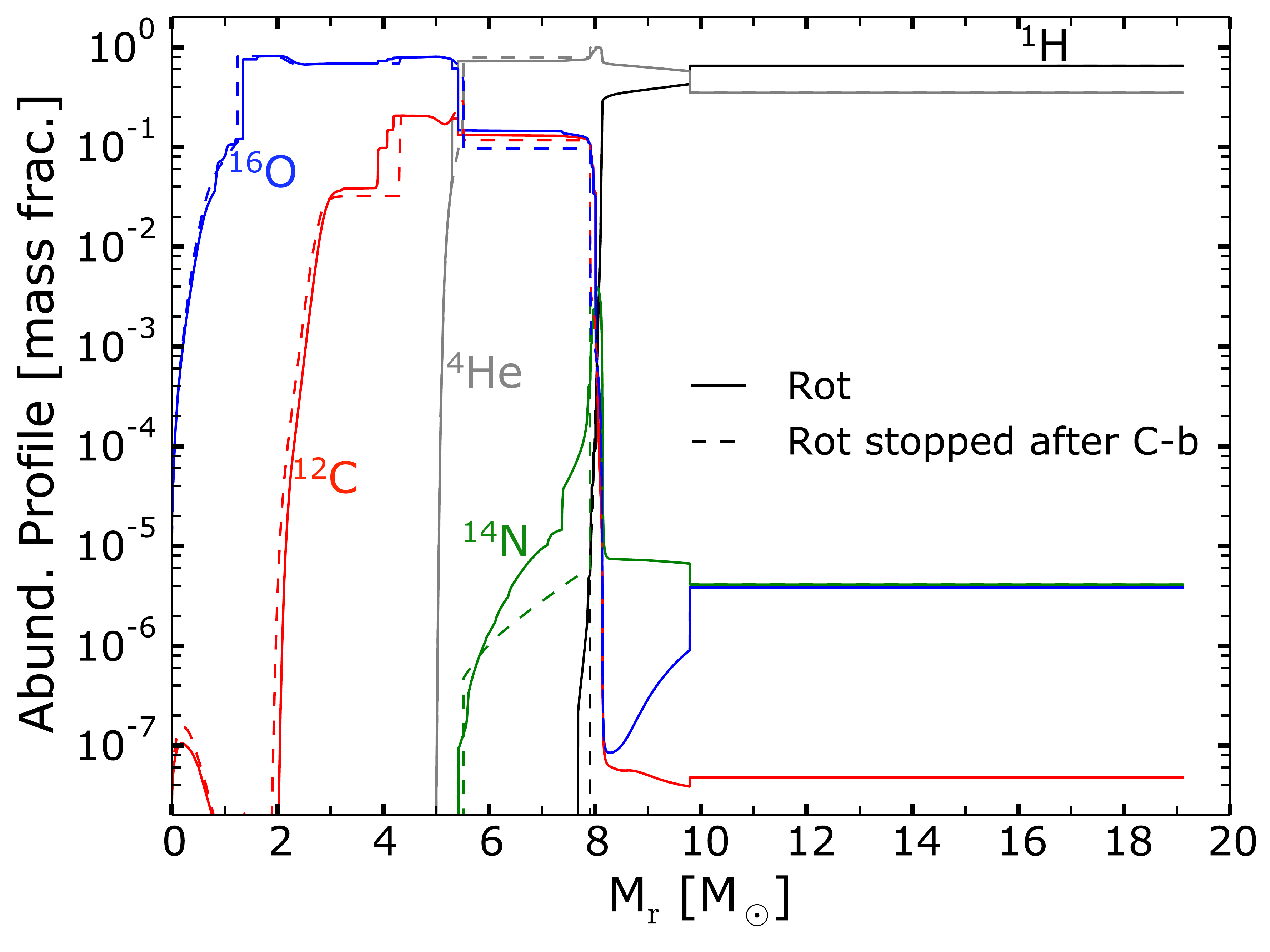}
   \caption{Abundance profiles of the 20s7 model at the end of the oxygen-burning phase when the rotation is taken into account until the end (solid line) and when the rotation is stopped at the end of the carbon-burning phase (dashed lines).}
              \label{stoprot} 
    \end{figure}

The evolutionary tracks of the six models of Table \ref{table:1} are shown in Fig. \ref{HRD}. They are computed until the end of the central silicon-burning phase, when the mass fraction of $^{28}$Si in the core is less than $10^{-8}$. We need a realistic pre-supernova structure of the star in order to obtain a reliable chemical composition of the supernova ejecta. As a consequence, reaching advanced stages of the evolution is an important point for the models presented in this work.  The effects of rotation are taken into account in the rotating models only until the end of the carbon-burning phase, which saves a lot of computational time and leads to only very small differences in the abundance profiles (Fig. \ref{stoprot}). This is mainly due to the fact that the duration of the last stages is short ($\sim 1 - 300$ days, c.f. Table \ref{table:1}) compared to the timescale of rotational mixing. Also, for most of the elements considered here, the explosive nucleosynthesis will have little impact. In that case, the final structures of the present models give a reasonable view of the chemical composition of the supernova ejecta.

\subsection{The chemical composition of the ejecta}\label{sec:3}

The source-star ejecta can be decomposed into wind- and supernova ejecta. The section below explains how these two types of ejecta are computed and combined together.

\subsubsection{The chemical composition of the wind}

As an example, let us consider the isotope $^{12}$C. First, we express the total ejected mass of $^{12}$C in the wind as a function of time $t$:
  \begin{equation}
  \label{mwind}
      M^\text{W}_{^{12}\text{C}}(t) = \int_\text{0}^{t} \dot{M}(t') X_{^{12}\text{C}}(t') \, \text{d}t',
   \end{equation}   
where $\dot{M}(t')$ is the mass-loss rate computed at time $t'$ and $X_{^{12}\text{C}}(t')$ the surface mass fraction of $^{12}$C at time $t'$. The integrated mass fraction of $^{12}$C in the wind as a function of time $t$ is then
     \begin{equation}
      X^{W}_{^{12}\text{C}}(t) = \frac{M^\text{W}_{^{12}\text{C}}(t)}{M^\text{W}(t)},
   \end{equation}   
with $M^\text{W}(t)$ the total mass of wind ejected after a time $t$. It is important to note that $X^\text{W}_{^{12}C}(t)$ is the mass fraction of $^{12}$C in the whole wind ejected from the ZAMS ($t=0$) to the time $t$. We have supposed here that the wind is homogeneously mixed.

\subsubsection{The chemical composition of the supernova}

The ejected mass of $^{12}$C in the supernova as a function of the mass cut, $M_\text{cut}$ , is
\begin{equation}
M^\text{SN}_{^{12}\text{C}}(M_\text{cut}) = \int_{M_\text{cut}}^{M_\text{fin}} X_{^{12}\text{C}}(M_r) \, \mathrm{d}M_r,
\end{equation}  
with $M_\text{fin}$ the mass of the star at the end of the evolution (see Table \ref{table:1}) and $X^\text{SN}_{^{12}C} (M_\text{r})$ the mass fraction of $^{12}$C, at the lagrangian coordinate $M_\text{r}$, at the end of evolution. $M^\text{SN}_{^{12}\text{C}}(M_\text{cut})$ corresponds to the mass of $^{12}$C in the part of the star that is expelled (layers between $M_\text{cut}$ and $M_\text{fin}$).

\subsubsection{The chemical composition in the wind and supernova combined}

To obtain the chemical composition in the wind and supernova ejecta combined, we add the material ejected through the wind to the material ejected through the supernova. The mass fraction of $^{12}$C in the ejected material as a function of  $M_\text{cut}$ is finally:
  \begin{equation}
  \label{mwindsn}
      X^\text{W+SN}_{^{12}\text{C}}(M_\text{cut}) = \frac{M^\text{W}_{^{12}\text{C}}(\tau_\text{life}) + M^\text{SN}_{^{12}\text{C}}(M_\text{cut})}{M^\text{W}(\tau_\text{life}) + M^\text{SN}(M_\text{cut})}
   ,\end{equation} 
 where $M^\text{W}_{^{12}\text{C}}(\tau_\text{life})$ is computed according to Eq.~\ref{mwind}, $\tau_\text{life}$ is the lifetime of the star (see Table \ref{table:1}), and $M^\text{W}(\tau_\text{life}) + M^\text{SN}(M_\text{cut})$ represents the total ejected mass.

\section{\boldmath The [C/N] $-\ ^{12}\text{C}/^{13}\text{C}$ puzzle}
\label{sec:3}

In the present section, the models of Table \ref{table:1} are discussed. We mainly focus on the [C/N] and $^{12}$C/$^{13}$C ratios, that provide interesting constraints on the possible CEMP-no source stars.

\subsection{Non-rotating models}
\label{20case}

   \begin{figure}
   \centering
       \includegraphics[scale=0.25]{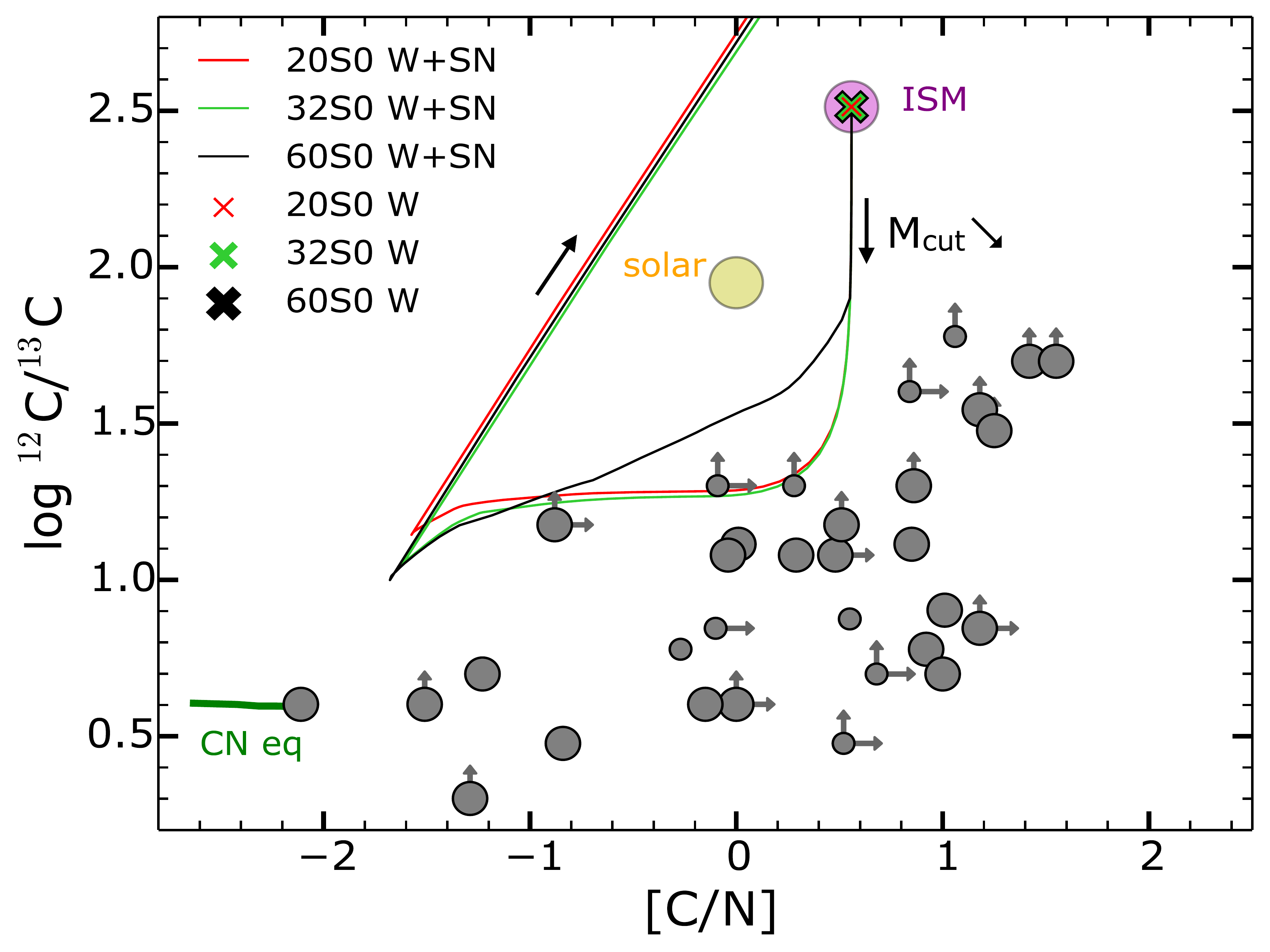}
   \caption{[C/N] vs. $\log(^{12}\text{C}/^{13}\text{C})$ diagram. Grey circles are ratios observed at the surface of CEMP-no stars with [Fe/H] $<-2.5$, [C/Fe] $>0.7$ and [Ba/Fe] $<1$ \citep{christlieb04, beers07, johnson07, lai08,masseron10, allen12,norris13, spite13, yong13, cohen13, roederer14a, hansen15}. Small circles are MS stars or subgiants while large circles are bright giants. The arrows indicate that only lower limits are deduced from spectroscopy. The yellow and purple circles represent the solar ratios and the ratio in an $\alpha$-enhanced ISM, respectively. The tracks represent the integrated ratios as more and more layers of the final structure are ejected and added to the wind (W+SN, effect of the mass cut, see Eq. \ref{mwindsn}) for the $20$, $32$, and $60\,M_\odot$ non-rotating models. The crosses show the ratios in the wind (W) at the end of silicon-burning (the crosses are superimposed). The thick green lines labelled `CN eq' represent the ratios obtained in a single zone at CN-equilibrium for 30 $<T<$ 80 MK.
}
\label{CNCC_norot}
    \end{figure}

Fig. \ref{CNCC_norot} shows the evolution in the [C/N] versus $\log(^{12}\text{C}/^{13}\text{C})$ diagram of the non-rotating models.  Since there is no rotational mixing operating inside these models, the surface is not enriched in $^{13}$C and $^{14}$N, so the ratios in the material ejected at the end of the evolution (crosses) are the same as the ratios in the initial ISM. The three crosses are thus superimposed, so that only the black one is visible. The red, green, and black lines show the effect of the mass cut (see Eq. \ref{mwindsn}). When varying $M_\text{cut}$ inward, we reach hotter and hotter regions where the CN cycle has operated so that [C/N] and $^{12}$C/$^{13}$C in the ejecta get closer to the CN equilibrium line. We note that even when the whole H-burning shell has been expelled, the ejecta is a mix between the initial ISM and the CN-processed material, so the CN equilibrium ratios are never reached. Expelling then deeper layers (where He has burnt and $^{13}$C and $^{14}$N have been completely depleted) strongly raises the two ratios (steep rise of the lines in Fig. \ref{CNCC_norot}).

   \begin{figure*}
   \centering
   \begin{minipage}[c]{.49\linewidth}
       \includegraphics[scale=0.25]{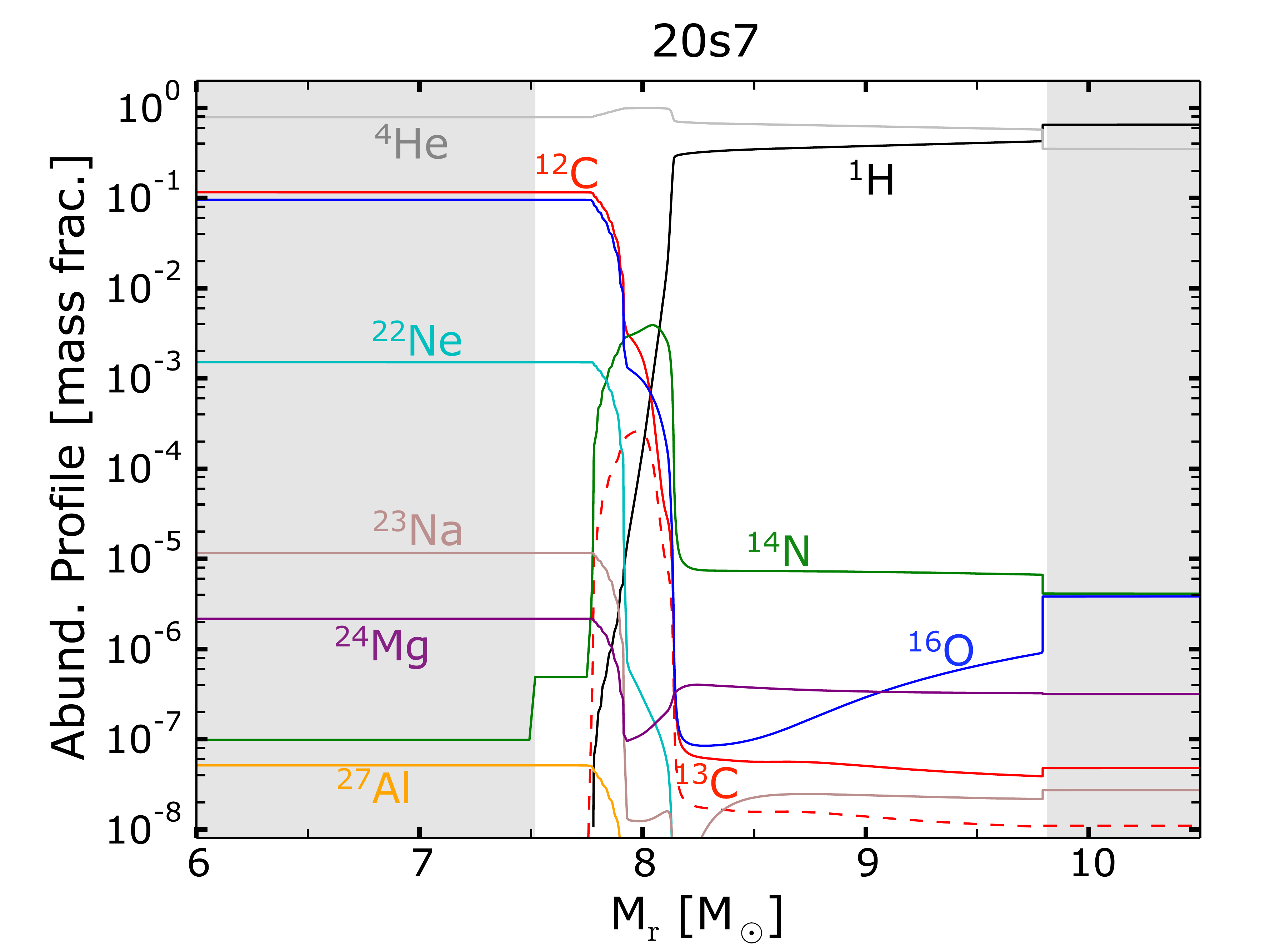}
   \end{minipage} \hfill
   \begin{minipage}[c]{.49\linewidth}
      \includegraphics[scale=0.25]{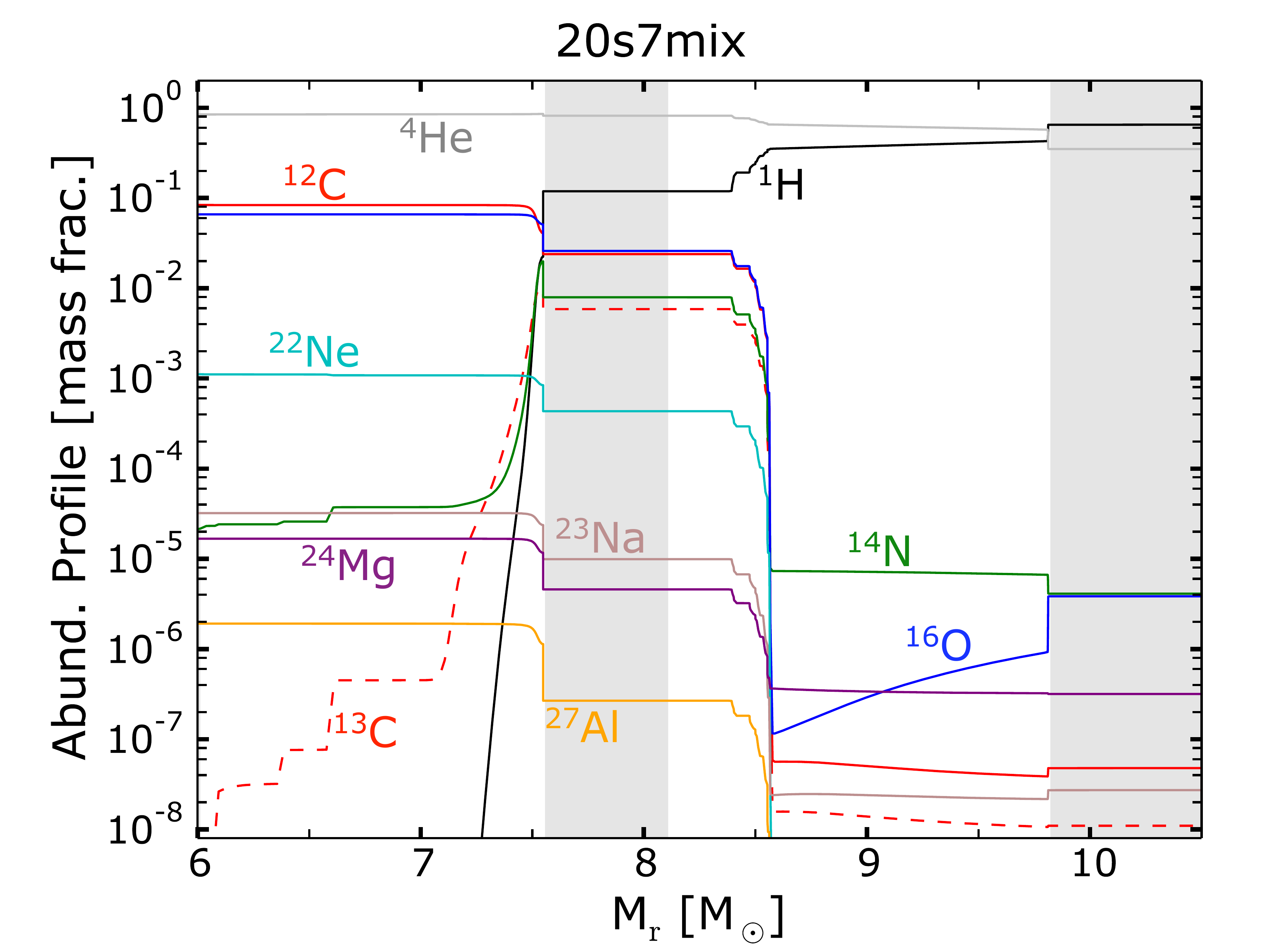}
   \end{minipage}
   \caption{Abundance profiles inside the 20s7 model at the end of the carbon-burning with no late mixing between the H- and He-burning shells (left) and with late mixing (right). Shaded areas show the convective zones.
}
\label{abundprof}
    \end{figure*}

\subsection{Rotating models}
The left panel of Fig. \ref{abundprof} shows the final abundance profile of the 20s7 model. The convective zone below $M_r \sim 7.5\,M_\odot$ corresponds to the He-burning shell. During the core He-burning phase and after, the products of He-burning (mainly $^{12}$C and $^{16}$O) diffuse from the He-burning region to the H-burning shell. This boosts the CNO cycle in the shell and creates primary $^{13}$C and $^{14}$N (hence the bump around mass coordinate $8\,M_\odot$). 

\begin{figure}
   \centering
      \includegraphics[scale=0.25]{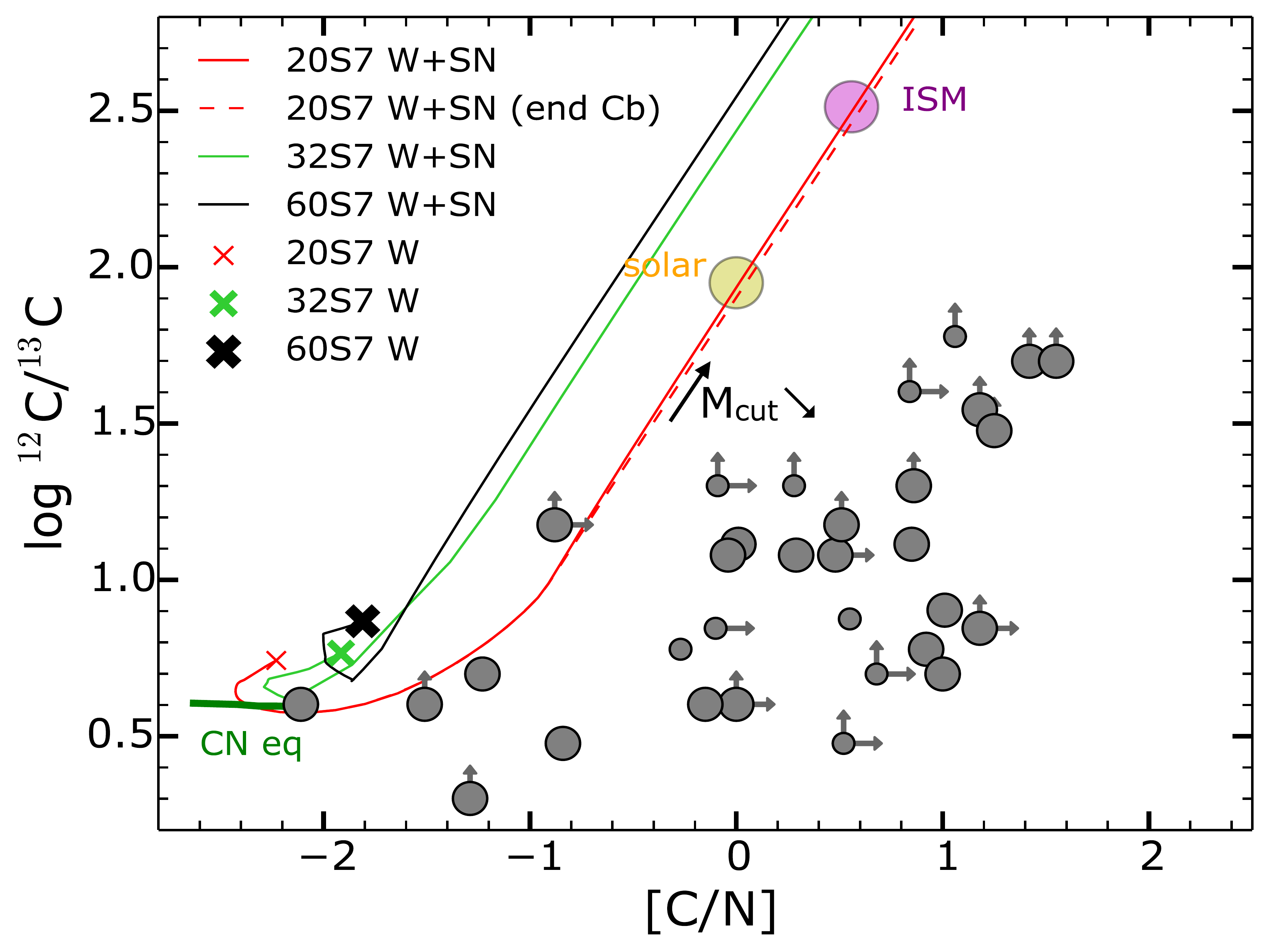}
   \caption{Same as Fig. \ref{CNCC_norot} but for rotating models. The red dashed line corresponds to the case where the evolution is stopped at the end of carbon burning instead of silicon burning for the 20s7 model. The chemical composition of the wind at the end of carbon burning and at the end of silicon burning is the same (red cross).}
\label{CNCC_rot}
    \end{figure}

\begin{figure}
   \centering
       \includegraphics[scale=0.49]{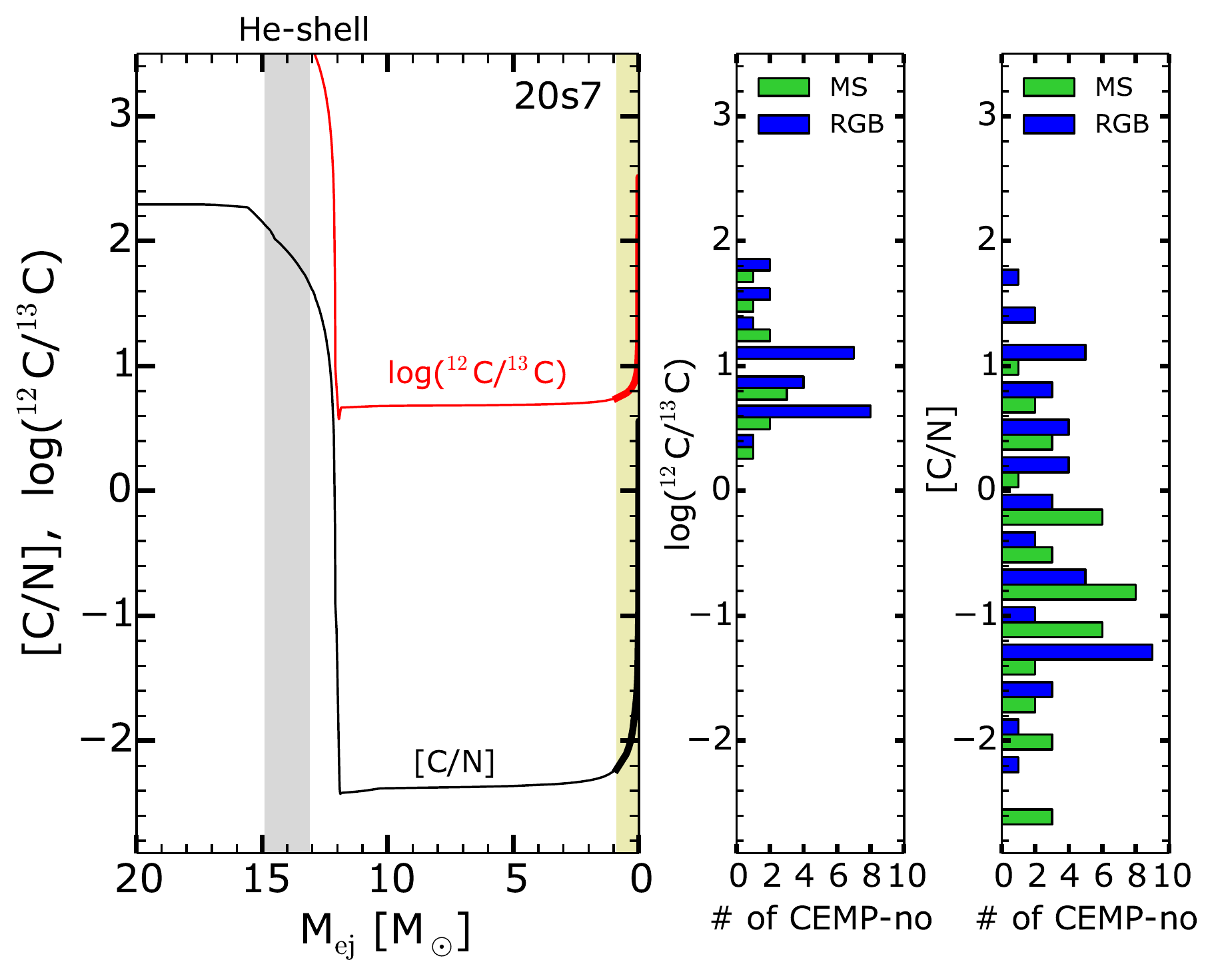}
   \caption{\textit{Left: }[C/N] and $\log(^{12}\text{C}/^{13}\text{C})$ in the ejecta as a function of the ejected mass for the 20s7 model. The thick lines in the yellow zone represent the integrated ratios in the wind (Eq. \ref{mwind}) as a function of $M_\text{ej}^\text{W} = \int_\text{0}^\text{t} \dot{M}(t') \, \mathrm{d}t'$. The thin lines show the integrated ratios in the ejecta (Eq. \ref{mwindsn}) as a function of $M_\text{ej}^\text{W+SN} = M_\text{ini} - M_\text{cut}$. The grey shaded area shows the location of the He-burning shell (where the energy produced by He-burning $\epsilon_\text{He}>10^3$ erg g$^{-1}$ s$^{-1}$) at the end of the evolution. \textit{Middle} and \textit{right}: distribution of observed $\log$($^{12}$C/$^{13}$C) and [C/N] for the CEMP-no stars with [Fe/H] $<-2.5$, [C/Fe] $>0.7$ and [Ba/Fe] $<1$ \citep{cohen04, honda04, christlieb04, sivarani06, beers07, johnson07, lai08, masseron10, caffau11, allen12, norris12, norris13, yong13, cohen13, spite13, placco14a,roederer14c,roederer14a, hansen14, hansen15}. The green histogram represents the MS stars or subgiants close to the turnoff with $T_\text{eff} > 5500$ K and $\log g \geq 3.25$, as by \cite{norris13}. The blue histogram shows the other stars.
}
\label{rapM}
    \end{figure}

Fig. \ref{CNCC_rot} shows that at the end of the evolution, the [C/N] and $\log(^{12}\text{C}/^{13}\text{C})$ ratios in the wind (crosses) are very close to characteristic CNO-equilibrium values (green line labelled `CN eq'). This is because during the evolution, the surface of the star is enriched in products of the CNO cycle brought from inner regions to the surface thanks to rotational mixing. This modified composition is then expelled in the wind. This point is also well illustrated by the left panel of Fig. \ref{rapM}, which focuses on the 20s7 model. It shows [C/N] and $\log(^{12}\text{C}/^{13}\text{C})$ ratios in the ejecta as a function of the ejected mass. The yellow area represents the wind. The thick lines are the integrated ratios in the wind as evolution proceeds, computed according to Eq. \ref{mwind}. [C/N] and $\log(^{12}\text{C}/^{13}\text{C})$ decrease in the wind, reaching $-2.2$ and $0.7$ at the end of evolution. 

The thin lines in Fig. \ref{rapM} show the effect of the mass cut. When only the outer layers are expelled, [C/N] and $^{12}\text{C}/^{13}\text{C}$ do not change because only the hydrogen-rich envelope is ejected in this case; this is a CN-processed region, having a similar composition to the wind. As we dig deeper into the source star, we reach a region processed by He-burning (at $M_\text{ej} \sim 12\,M_\odot$) so that [C/N] and $^{12}\text{C}/^{13}\text{C}$ increase dramatically in the ejecta. As a complement, the red line in Fig. \ref{CNCC_rot} shows how [C/N] and $^{12}$C/$^{13}$C are linked together when varying the mass cut for the 20s7 model. The 32s7 and 60s7 models are also shown and behave similarly to the 20s7 model.

\subsection{Can a low $^{12}$C/$^{13}$C with a high [C/N] be reproduced by models?}

The middle and right panels of Fig. \ref{rapM} show the observed $^{12}$C/$^{13}$C and [C/N] distribution of CEMP stars with\footnote{In this plot, some CEMP-no stars have a different [Fe/H] than the one of the models (about $-3.8$). We think that this is not a problem since we are looking at isotope ratios; the CNO equilibrium value of $^{12}$C/$^{13}$C does not vary with metallicity for instance.} [Fe/H] $<-2.5$, [C/Fe] $>0.7$ and [Ba/Fe] $<1$. The condition on [Ba/Fe] rules out the CEMP stars significantly enriched in barium, generally classified as CEMP-s, -r/s or -r stars. While the distribution of observed $\log(^{12}\text{C}/^{13}\text{C})$ peaks close to the CN-equilibrium value of $\sim 0.6$, the distribution of [C/N] spans a wide range, extending largely above the CN-equilibrium value of $\sim -2.3$. To reproduce the whole [C/N] range, one needs to consider material coming from layers belonging both to the outer layers of the source stars and from deeper layers, having been processed by He-burning. In contrast, to reproduce the observed variations of $\log(^{12}\text{C}/^{13}\text{C})$, no layers that have been processed by He-burning should be involved. We note that although some stars have a measured $^{12}$C/$^{13}$C close to CN-equilibrium \citep[e.g. CS22945-017 and HE0007-1832, which are MS CEMP-no with $\log(^{12}\text{C}/^{13}\text{C})$ $= 0.8$ and $0.9$ respectively][]{masseron10, cohen04}, some others have only a lower limit, requiring some caution regarding the previous statements.  Arrows mark these limits in Figs. \ref{CNCC_norot} and \ref{CNCC_rot}. The tracks on these two figures clearly show that the ejecta of the six source-star models are unable to provide a solution matching the observed bulk of CEMP-no stars. We remark however that although globally out of the observed range, these models might explain some CEMP-no stars. This is discussed in more detail in Sect. \ref{sec:6} and \ref{sec:7}.

An important point here is that, in any case, only the outer layers of the source stars should be ejected, otherwise the $^{12}$C/$^{13}$C ratio in the ejecta is largely above the bulk of observed values. This being said, the difficulty remains of having a high C/N at the same time in the ejecta.

At this point, we think that two important points need to be addressed regarding $^{12}$C/$^{13}$C: (1) Can the source star ejecta be diluted with the ISM in such a way that the $^{12}$C/$^{13}$C ratio is reduced? 
(2) Can the CEMP-no stars themselves have modified their surface composition?

\subsection{Dilution with the interstellar medium}

Let us consider that the source star could have expelled a material with a high $^{12}$C/$^{13}$C ratio that would afterwards be diluted with the ISM, having a lower $^{12}$C/$^{13}$C ratio. The values of the $\log(^{12}\text{C}/^{13}\text{C})$ ratio for a solar and $\alpha$-enhanced ISM are about 2.5 and 2 respectively (see purple and yellow circles in Fig. \ref{CNCC_rot}). The dilution of a material with a high $^{12}$C/$^{13}$C ratio coming from the source star with the ISM can lead to  $\log(^{12}\text{C}/^{13}\text{C})$ values of 2.5 and 2 at best, respectively. Relying on such an ISM, the dilution  cannot likely be a solution. Of course, the composition of the ISM at such low metallicity is barely known and might be different. With chemical evolution models, \cite{chiappini08} have shown that if fast rotators were dominant in the early universe, the $\log(^{12}\text{C}/^{13}\text{C})$ ratio would lay between $1.5$ and $2.5$ in the ISM. This is still above the bulk of the $\log(^{12}\text{C}/^{13}\text{C})$ distribution of CEMP-no stars. Considering an even lower $^{12}$C/$^{13}$C ratio in the ISM might be a solution. However, in that case, we think that this would simply push the problem further: where do such low $^{12}$C/$^{13}$C ratios come from?

\subsection{In situ modification of CEMP-no stars}

The CEMP-no stars could have changed their surface abundances because of internal processes. In that case, the comparison between their surface abundances and the material ejected by the source star is more difficult. The two main processes known are (1) atomic diffusion and (2) the first dredge up. The atomic diffusion comprises different processes like gravitational settling and radiative acceleration. Since CEMP-no stars are old, these processes may have had time to change their surface composition. However, the $^{12}$C/$^{13}$C ratio is barely affected by gravitational settling or by radiative acceleration since the two isotopes have similar weights and electronic transitions. Atomic diffusion in a more general context is discussed in Sect. \ref{sec:7}.

The first dredge up, occurring after the main sequence, brings internal material up to the surface of the CEMP-no star. As this material comes from hotter regions, where the CN cycle is likely operating, it is enriched in $^{14}$N and $^{13}$C and depleted in  $^{12}$C. As a consequence, the dredge up is expected to decrease the surface [C/N] and $^{12}$C/$^{13}$C ratios. We note that the ON branch is likely not activated because of the overly low temperature in these low-mass stars. It is particularly interesting to compare the [C/N] and $^{12}$C/$^{13}$C ratios of the MS sample with the RGB sample. A possible guess based on the previous discussion is that the RGB stars should present lower [C/N] and $^{12}$C/$^{13}$C ratios than MS stars due to the effect of the first dredge up. The green histograms in Fig. \ref{rapM} show MS stars or subgiants close to the turnoff. The blue histograms show the other stars, classified as RGB.
We see that for both [C/N] and $^{12}$C/$^{13}$C, the MS sample covers the same range of values as the RGB sample. This shows that the effect of the first dredge up is probably small compared to the effect of the observed dispersion of the [C/N] and $^{12}$C/$^{13}$C ratios. Should it be a strong effect, we would see a clear separation between MS and RGB stars.

This discussion shows that most likely the abundances that are observed at the surface of CEMP-no stars do not result from in situ processes but reflect indeed the composition of the cloud from which they formed.

   \begin{figure}
   \centering
      \includegraphics[scale=0.47]{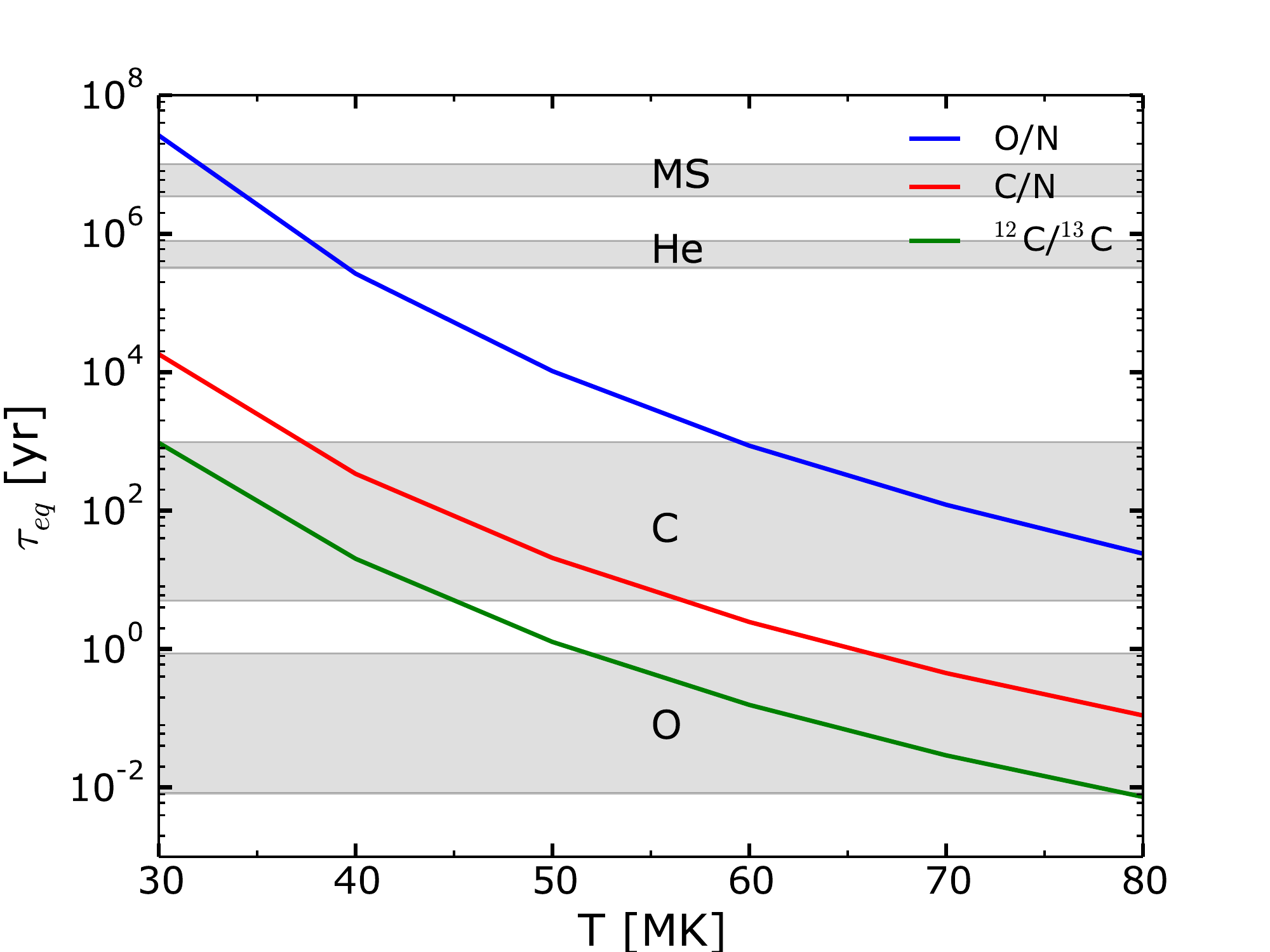}
   \caption{Equilibrium timescales of O/N, C/N, and $^{12}$C/$^{13}$C as a function of the temperature. We used a one zone model at density $\rho =$ 1 g cm$^{-3}$. The shaded areas show the ranges of duration for the various burning stages (main sequence, He-, C-, and O-burning) of the models presented in Table \ref{table:1}. }
              \label{eqtime} 
    \end{figure}

   \begin{figure*}
   \centering
   \begin{minipage}[c]{.49\linewidth}
       \includegraphics[scale=0.25]{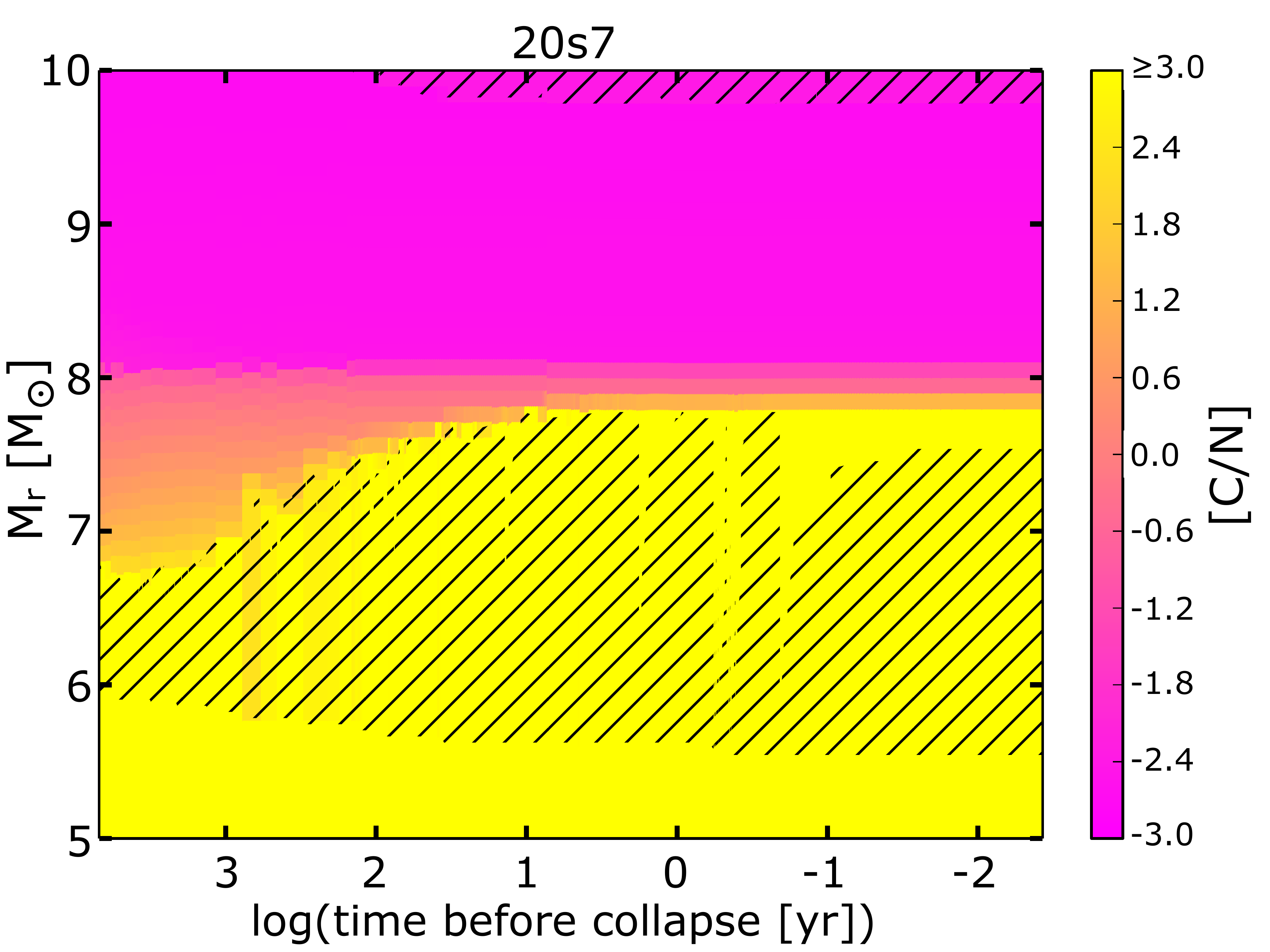}
   \end{minipage} \hfill
   \begin{minipage}[c]{.49\linewidth}
      \includegraphics[scale=0.25]{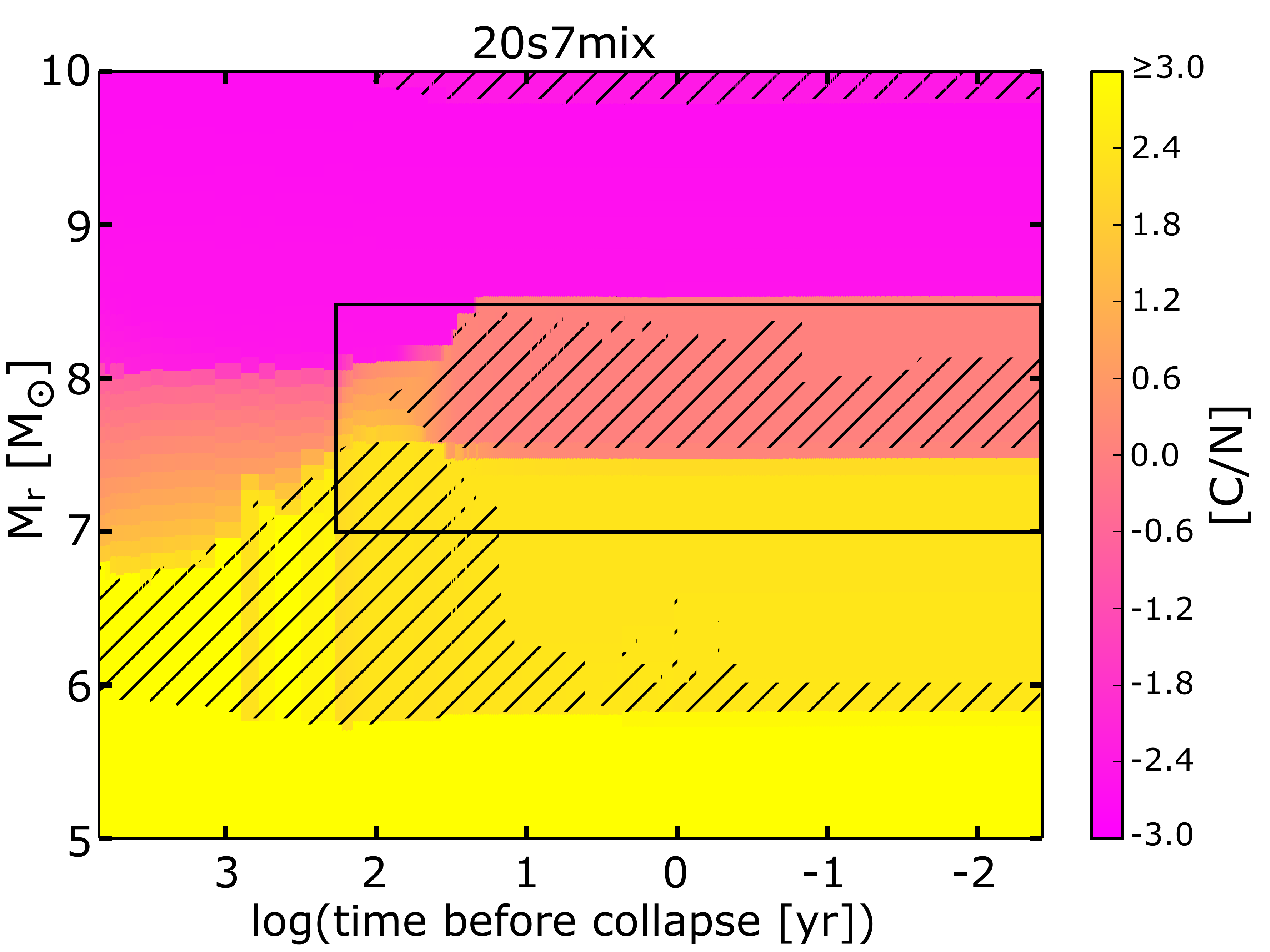}
   \end{minipage}
   \caption{Evolution of the structure as a function of the time left until the core collapse (Kippenhahn diagram) for the 20s7 model without (left) and with (right) late mixing. The colour-scale shows the value of [C/N] inside the star. Hatched areas represent the convective zones. The black frame on the right panel indicates where and when the mixing is enhanced.}
\label{kip}
    \end{figure*}

\subsection{A partially CN-processed material?}

We have seen above that the present source star models cannot reproduce the CEMP-no stars showing simultaneously values of the $^{12}$C/$^{13}$C ratio typical of the CNO equilibrium and C/N ratios above the CNO equilibrium value.
We propose here a possible solution to the C/N $-$ $^{12}\text{C}/^{13}\text{C}$ puzzle.

Figure~\ref{eqtime} shows the equilibrium timescales of the C/N (red) and $^{12}$C/$^{13}$C (green) ratios as a function of the temperature. O/N is also  shown for comparison. We used a one-zone model at constant temperature and density. We considered temperatures ranging from 30 to 80 MK. The timescales are taken when the ratios in the zone are equal to 99$\%$ of their equilibrium value. When the temperature increases, the ratios reach the equilibrium value more quickly. From 30 to 80 MK, the timescales decrease by $\sim 4-5$ orders of magnitude. Also, whatever the temperature, $^{12}$C/$^{13}$C reaches the equilibrium approximately ten times faster than C/N, and C/N reaches the
equilibrium approximately $100-1000$ times faster than O/N.

In a massive and low-metallicity source-star model, the zone where the H is burning (core or shell) can be convective. In that case, the material is assumed to be mixed instantaneously. Each mesh point in this H-burning zone has a different temperature (between $\sim 30$ and $\sim 80$ MK), hence a different equilibrium timescale.
However, since the equilibrium timescales change monotonically with the temperature, the global equilibrium timescales in the convective H-burning zone are bound between the timescales at 30 and 80 MK. Also, the relative difference between these three global equilibrium timescales remains the same as the difference shown in Fig. \ref{eqtime}.

Let us now consider that some $^{12}$C is injected into the convective H-burning shell which is initially at CNO equilibrium. We consider that the global equilibrium timescales of this shell correspond to the timescales at 40 MK ($\sim 30$ yr for $^{12}$C/$^{13}$C, $\sim 300$ yr for C/N, see Fig.~\ref{eqtime}).
After a length of time of between $\sim 30$ and $\sim 300$ yr, the $^{12}$C/$^{13}$C will be at equilibrium while the C/N will not.
Thus we see that if the injection of $^{12}$C occurs less than $\sim 300$ yr before the core collapse, we would have some part of the star with a chemical composition potentially resolving the C/N $-$ 12C/13 puzzle.
From Fig. \ref{eqtime}, we can deduce that this injection should occur after the core He-burning phase, during the C or O-burning phase (in our models, C burning lasts for 10 $-$ 1000 yr for instance). The injection of $^{12}$C can occur if a mixing event happens between the H- and He-burning regions.
The mixing event should be strong enough so that sufficient $^{12}$C is injected. If not, the CNO cycle quickly returns to equilibrium.

Mixing naturally happens in our rotating models but remains mild enough so that C/N and $^{12}$C/$^{13}$C in the hydrogen shell are at CNO equilibrium at the end of evolution. A stronger injection of $^{12}$C in the hydrogen shell is needed to boost the CNO cycle more and leave an excess of $^{12}$C with respect to $^{14}$N at the end of evolution.
We have explored the consequences of this idea by considering such late mixing episodes in our stellar models.

\section{A late and strong mixing process in source stars}
\label{sec:4}

\subsection{The recipe}\label{sec:41}

For the six models of Table \ref{table:1}, we have triggered a late mixing event $\sim$ 200 yr before the end of the core C burning phase. The end of this phase is defined such that X($^{12}$C)$_c$, the central mass fraction of $^{12}$C is equal to 10$^{-5}$.
From Table \ref{table:1}, we know that the duration of core C burning can be shorter than 200 yr. In such cases, the late mixing begins before the C starts to burn but still after the core He-burning phase. Late mixing is only operating in radiative zones and is triggered around the bottom of the H-burning shell. To model this late mixing process in rotating models, we multiply the $D_\text{shear}$ coefficient (Eq. \ref{dshearTZ}) by a factor of 100. Non-rotating models are also studied to see whether or not the effect of the late mixing alone might be sufficient to explain\ the abundances observed at the surface of CEMP-no stars. In those models, we set $D_\text{shear} = 10^{9}\,\text{cm\,s}^{-1}$ in the late mixing zone, a characteristic value found in rotating models with late mixing. These new models\footnotemark 
\footnotetext{These new models do not appear in Table~\ref{table:1}. They have the same properties as the models without late mixing, shown in Table~\ref{table:1}. Only the duration of the C-burning phase changes slightly.}
are identified with `mix' (20s0mix, 32s0mix...). 
Although modelled with a shear diffusion coefficient, we do not assume that the physical origin of the late mixing process is linked to the shear. Its possible physical origin is discussed in Sect. \ref{late}. Also, a parametric study of this late mixing is done in Sect. \ref{sec:5}.

   \begin{figure*}
   \centering
   \begin{minipage}[c]{.49\linewidth}
       \includegraphics[scale=0.25]{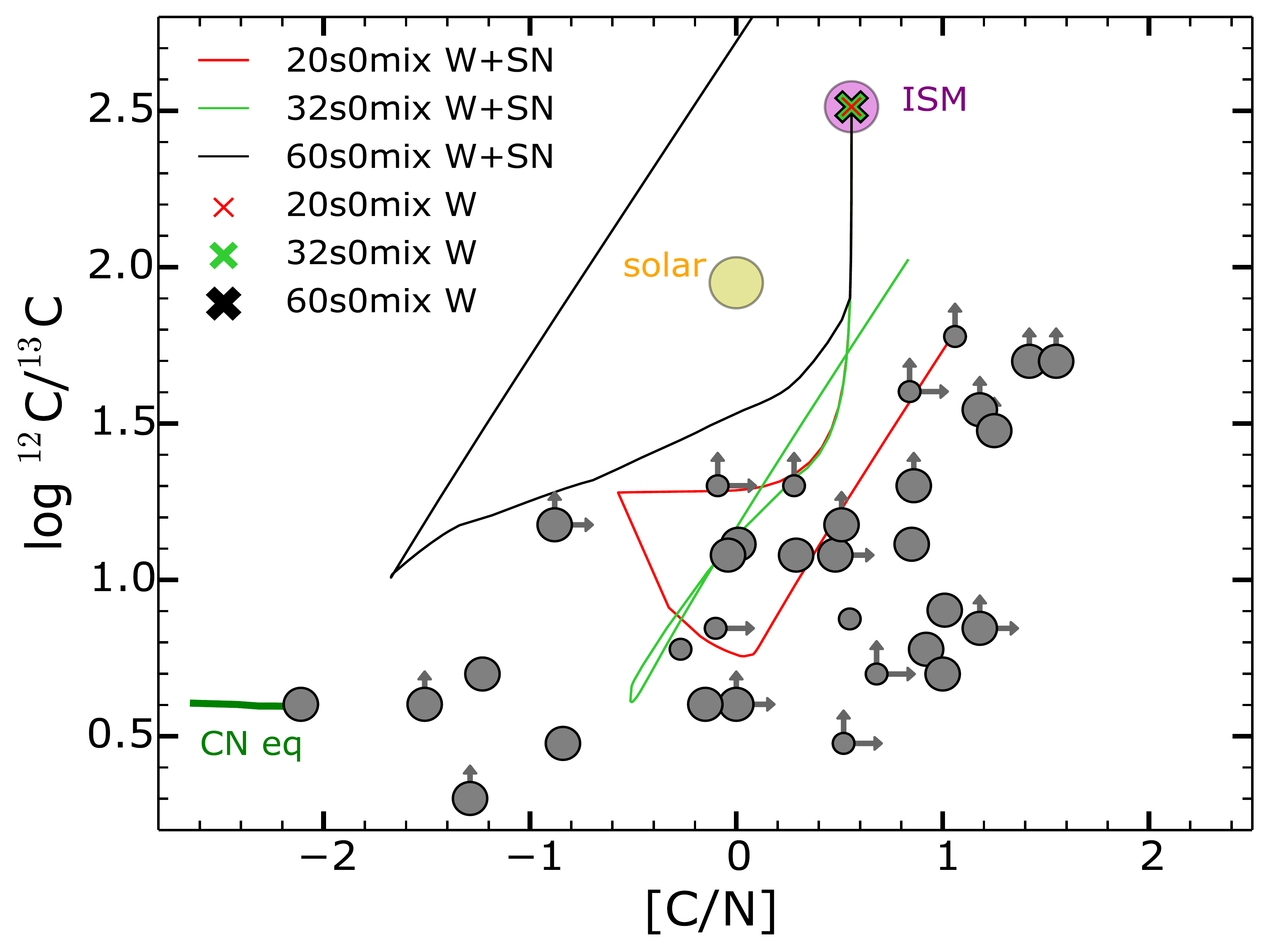}
   \end{minipage} \hfill
   \begin{minipage}[c]{.49\linewidth}
      \includegraphics[scale=0.25]{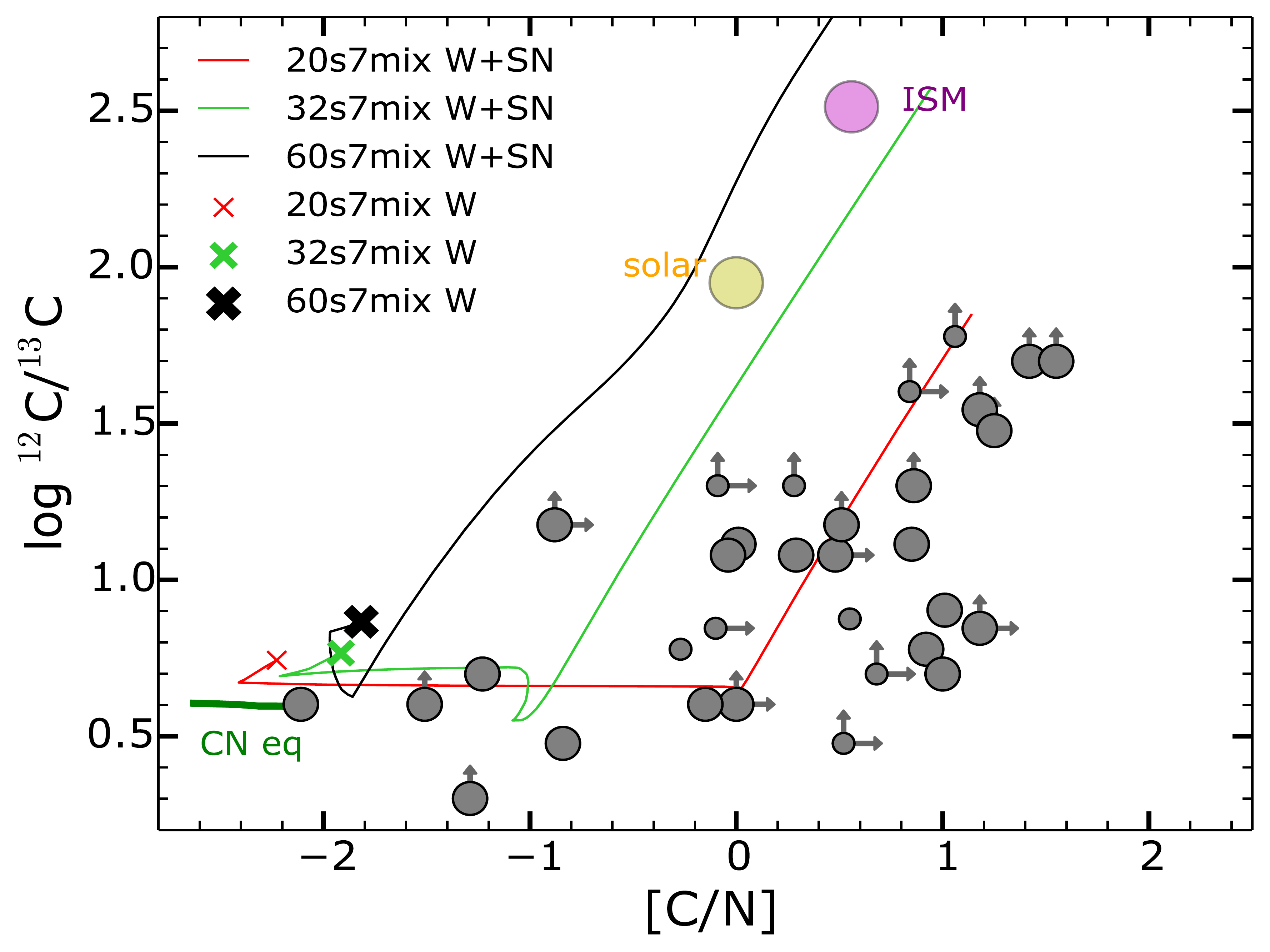}
   \end{minipage}
   \caption{As in Figs. \ref{CNCC_norot} and \ref{CNCC_rot} but for the models with late mixing (see Sect. \ref{sec:41} for explanation). The endpoint of the lines correspond to a mass cut located at the bottom of the He-burning shell. \textit{Left panel:} non-rotating models; \textit{right panel:} rotating models.}
\label{CNCCmix}
    \end{figure*}

\subsection{Late mixing in the 20s7 model}

Fig. \ref{kip} shows a Kippenhahn diagram of the 20s7 (left panel) and 20s7mix (right panel) models. Only the post core He-burning stage is shown, where there is both a He-burning and a H-burning shell. On the left panel, the convective zone between $\sim 6$ and $8\,M_\odot$ (hatched area) corresponds to the He shell. On the right panel, the lower convective zone corresponds to the He shell and the upper one, appearing at the abscissa $\sim 1.6$, corresponds to the H shell. The black frame shows the region in space and time where the mixing is enhanced.

Without late mixing, the mixing is mild enough so that not too much $^{12}$C diffuses from the He-burning shell to the H-burning shell. As a consequence, the CNO cycle is not very active and at the end of the evolution, we distinguish (1) a zone with a high [C/N] where He-burning has destroyed $^{14}$N (yellow zone) and (2) a zone with [C/N] at CNO-equilibrium (magenta zone).

With late mixing, more $^{12}$C enters the H shell, boosting the CNO cycle. The H shell becomes convective and more He-burning products are engulfed. The fresh C starts to be transformed into $^{13}$C and $^{14}$N in the H shell. However, the time remaining before the end of the evolution being short, the [C/N] equilibrium value of $\sim -2.3$ is not reached. The right panel of Fig. \ref{abundprof} shows the abundance profile of the 20s7mix model at the end of core C-burning. We clearly see the convective H shell with a lot of CNO elements, and where X(C)/X(N) $>1$ while X($^{12}$C)/X($^{13}$C) is at equilibrium, around 5. The right panel of Fig. \ref{kip} shows this intermediate zone where [C/N] is about 0 at the end of evolution. The late mixing process is then able, for the rotating 20 $M_\odot$ model, to build a zone partially processed by the CNO cycle in the source star, where C/N is high and $^{12}$C/$^{13}$C at equilibrium.

Fig. \ref{CNCCmix} shows the same results as Figs. \ref{CNCC_norot} and \ref{CNCC_rot} but for the models with late mixing. The 20s7mix model is represented on the right panel by the red track. Because of the partially CN-processed zone, there now exists a solution able to better reproduce the observed C/N $-$ $^{12}$C/$^{13}$C feature. The partially CN-processed zone for this model is characterised by $\log(^{12}\text{C}/^{13}\text{C})$ $\sim 0.6$ and [C/N] $\sim 0$, explaining the plateau going from [C/N] $\sim -2.5$ to $\sim 0$. For the models with late mixing, we stopped the evolution at the end of the core C-burning phase. We did not compute the very end of the evolution but this would lead to only very small changes in the abundance profiles of the outer layers, above the C-burning shell (see the red solid and dashed lines in Fig. \ref{CNCC_rot}).

\subsection{Late mixing in the other models}

The right panel of Fig. \ref{CNCCmix} shows that the 32s7mix (green) and 60s7mix (black) source-star models do not behave like the 20s7mix model. 
This is mainly because higher-mass models have a higher temperature in the convective H shell meaning that the CN cycle is faster. In this case, $^{12}$C is transformed more rapidly into $^{14}$N. The [C/N] ratio in the H-burning shell is then closer to the equilibrium value at the end of evolution.
This explains the different lengths of the plateau in the tracks shown on the right panel of Fig.~\ref{CNCCmix}. The longest plateau is for the 20s7mix model (red track) that has the lowest temperature in the H-burning shell. A low temperature implies that the CN cycle has not significantly operated. This gives a high [C/N] ratio at the end of the evolution. The 32s7mix plateau reaches lower [C/N] ratios (about $-1$) because of the higher rate of the CN cycle. In the 60s7mix model, there is no plateau; the zone is at CN equilibrium at the end of the evolution because of the even higher temperature in the H-burning shell. 

Triggering a late mixing in the non-rotating models also leads to this partially CNO-processed zone for the 20s0mix and 32s0mix models. For the 60s0mix model, the distance between the two shells is too large; the He-products ($^{12}$C and $^{16}$O) do not reach the H-burning shell.
The left panel of Fig. \ref{CNCCmix} shows that the 20s0mix and 32s0mix models might also yield a material able to better reproduce the observed distribution. Sect. \ref{sec:6} investigates how the `s0mix' and `s7mix' models can be discriminated by considering other chemical species.

An important point here is that the late mixing is more efficient in building a zone with a high C/N in the $20\,M_\odot$ source star than in more massive models, mainly because the temperature in the H-burning shell is lower in a $20\,M_\odot$ model, implying a slower pace of the CN cycle. The time window in which the late mixing process would give a partially CNO-processed material at the end of evolution is much longer for the $20\,M_\odot$ source star. This makes the $20\,M_\odot$ source stars better candidates for the late mixing scenario. We note, nevertheless, that if the late mixing occurs sufficiently late (later than $\sim 200$ yr before the end of the evolution), the material might be partially processed even in a $60\,M_\odot$ source star, where the CN cycle is faster. In this case however, the mixing event should be extremely strong so as to compensate for the short time available.

   \begin{figure*}
   \centering
   \begin{minipage}[c]{.49\linewidth}
       \includegraphics[scale=0.49]{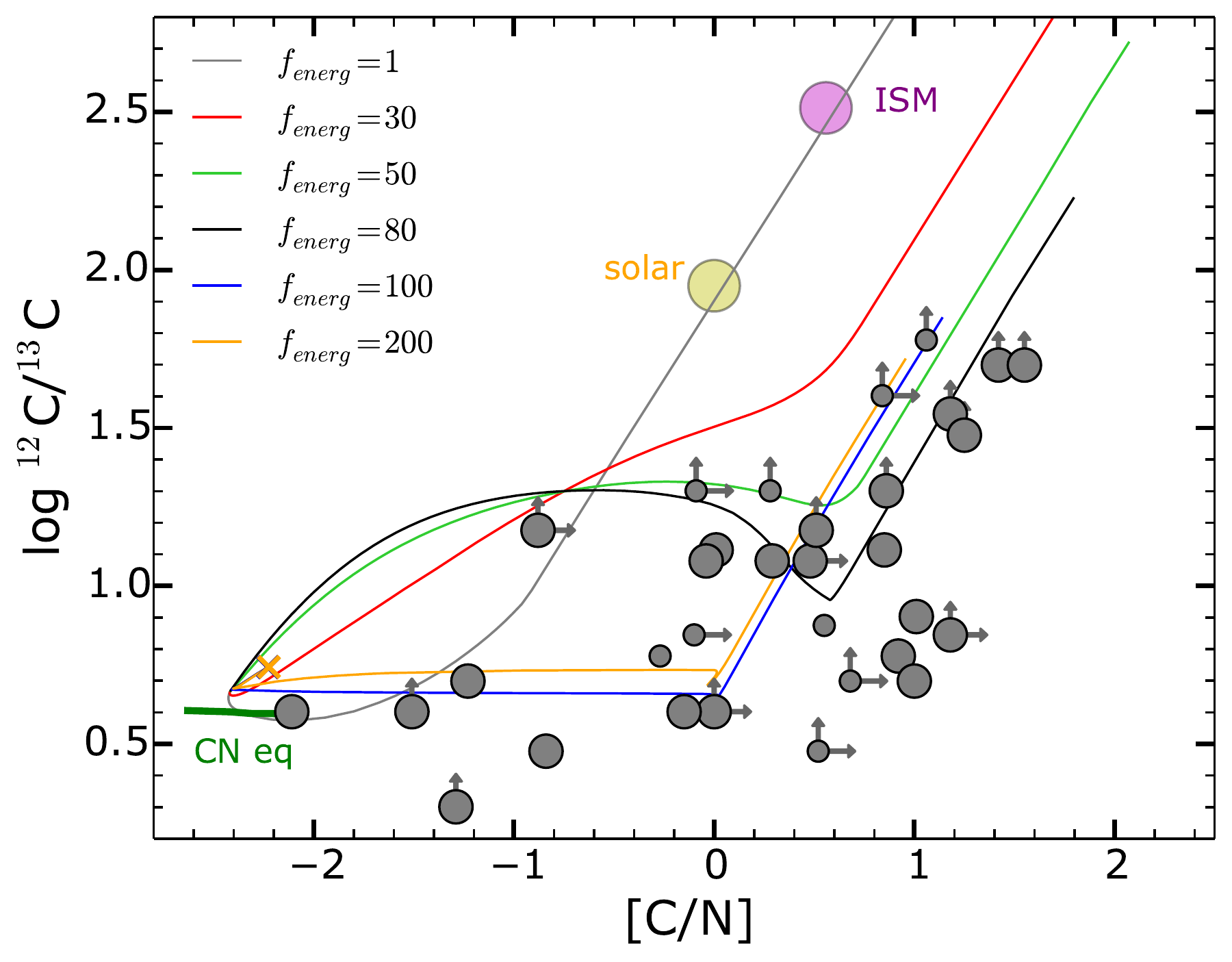}
   \end{minipage} \hfill
   \begin{minipage}[c]{.49\linewidth}
      \includegraphics[scale=0.49]{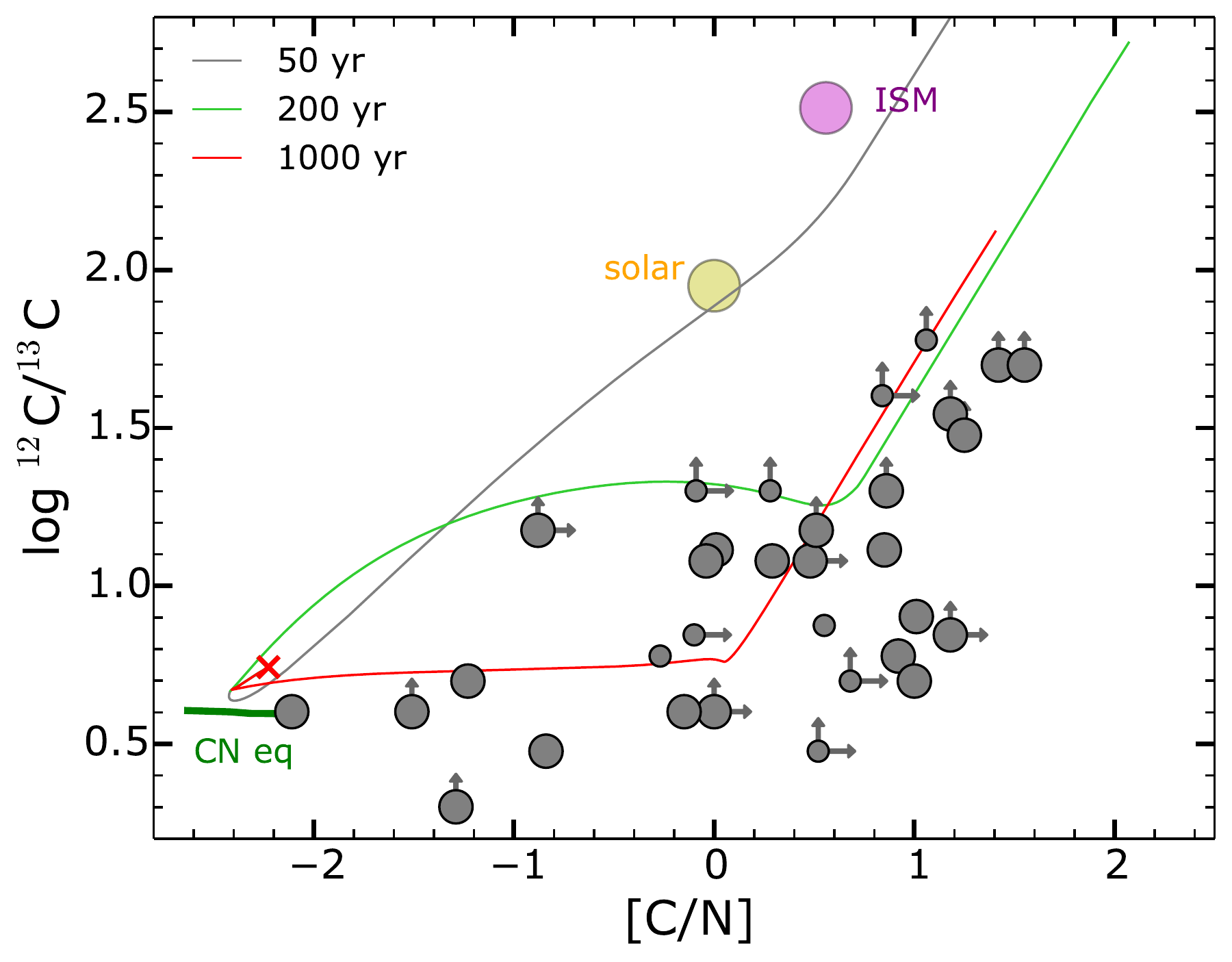}
   \end{minipage}
   \caption{As in Fig. \ref{CNCCmix} but when varying the $f_\text{energ}$ parameter (see Eq. \ref{dshearTZ}) for the 20s7mix model (left panel) and when triggering the late mixing at different times ($\sim 50$, $200$ and $1000$ yr before the end of C-burning), with $f_\text{energ}=50$ (right panel). The endpoint of the lines corresponds to a mass cut located at the bottom of the He-burning shell.}
\label{ftrig}
    \end{figure*}

\section{Varying the parameters of the late mixing event}
\label{sec:5}

Since the late mixing is artificially triggered, it is natural to wonder what happens if we vary the intensity of this process for instance, or if we trigger the late mixing at different times close to the end of the evolution. Changing the spatial extension of the late mixing zone is not investigated. We focus here on a rather small zone in the star, near the bottom of the hydrogen burning shell.

\subsection{Intensity of the late mixing}

Figure \ref{ftrig} ({\it left panel}) shows the effect of varying the intensity of the late mixing in the 20s7 model by changing the value for $f_\text{energ}$ (see Eq. \ref{dshearTZ}). The late mixing is triggered $\sim 200$ yr before the end of evolution, using $f_\text{energ} = 1$ (standard case), $30$, $50$, $80$, $100$ (case discussed in the previous section), and $200$. As $f_\text{energ}$ increases, more and more $^{12}$C enters into the H-burning shell. We see that the stronger the mixing (higher $f_\text{energ}$) the easier we reach regions of low $^{12}$C/$^{13}$C and non-equilibrium C/N ratios.

Let's focus on the shape of the $f_\text{energ} = 80$ curve (black); it shows that four zones exist in this source-star model:
\begin{enumerate}
\item A CN processed zone (external layers, represented by the orange cross). This zone is similar to the region above $M_\text{r} \sim 8.6\,M_\odot$ in the right panel of Fig. \ref{abundprof}\footnote{  Figure \ref{abundprof} actually corresponds to the $f_{energ} = 100$ model while we discuss here the $f_{energ} = 80$ model. Although slightly different, the abundance profiles of these two models show many similarities.}.
\item A zone where $^{12}$C is overabundant with respect to $^{13}$C and $^{14}$N (the curve rises).
\item A partially CN processed zone where $^{12}$C/$^{13}$C decreases towards equilibrium (or closeby) and C/N is high. This zone is similar to the region between $\sim 7.5$ and $\sim 8.5\,M_\odot$ in the right panel of Fig. \ref{abundprof}.
\item A region processed by He-burning that makes the curve rise dramatically (similar to the region below $M_\text{r} \sim 7.5\,M_\odot$ in the right panel of Fig. \ref{abundprof}).
\end{enumerate}

The bump of the black curve in this diagram is explained when we understand what happens in zones 2 and 3. Zone 2 is formed because some $^{12}$C diffuses into regions where $T_\text{6} \lesssim 30$ MK. At such a temperature, C/N and $^{12}$C/$^{13}$C reach CNO equilibrium after $\sim 10^4$ and $\sim 10^3$ yr, respectively (see Fig.~\ref{eqtime}). This is more than the remaining time before the end of evolution so that there is a carbon excess at the end. Zone 3 corresponds to deeper (and hotter) layers that have become convective once a sufficient amount of carbon has diffused and boosted the CNO cycle. In this zone, the CNO cycle is closer to equilibrium at the end of the evolution because of the higher temperature compared to zone 2.

For the $f_\text{energ} = 100$ and $200$ models (blue and orange curves), the zone 2 does not exist. Indeed, the convective zone that forms because of the energy released by the CNO cycle expands further out and grows in time, engulfing the zone 2. In the end, this induces sharp transitions between the CNO processed zone (orange cross), the partially CNO processed zone (around coordinates (0,0.6) for the orange curve) and the He-processed zone (dramatic rise of the curve).

\subsection{Time of the late mixing}

Figure \ref{ftrig} ({\it right panel}) shows the effect of varying the time at which the mixing is triggered. For this study we set $f_\text{energ}=50$. For a given $f_\text{energ}$, if the mixing is triggered too late, not enough $^{12}$C diffuses into the H shell and it does not create the partially CNO processed zone (see the grey line). If it is triggered $\sim 1000$ yr before the end, the partially CNO processed zone is created. We see that this model (red curve) is behaving like the models with a higher $f_\text{energ}$ in the left panel (blue and orange lines), illustrating that a stronger intensity for the mixing or a longer time for it to occur can have similar results.

To summarise what precedes:
\begin{itemize}
\item The late mixing process has to occur sufficiently late in the evolution otherwise [C/N] have enough time to go back to CN equilibrium (c.f. Fig. \ref{eqtime});
\item the partially-processed zone is more likely built in a $20\,M_\odot$ source star than in a $60\,M_\odot$;
\item if the mixing event is triggered in a $20\,M_\odot$ source star $\sim 200$ yr before the end of the evolution, the mixing has to be enhanced by at least a factor of 50 for the partially CNO-processed zone to be created;
\item if the mixing event is triggered earlier, the same behaviour is found provided the mixing has been reduced. This would be the opposite if the mixing event was triggered later (e.g. during O-burning); the mixing has to be increased to compensate for the shorter available time before evolution ends.
\end{itemize}

For what follows, we select the cases (for each model) where the late mixing is triggered $\sim 200$ yr before evolution ends and with $f_\text{energ} = 100$.

\section{Nucleosynthetic signatures of the source-star models}
\label{sec:6}

In this section, we highlight nucleosynthetic differences between the four categories of model computed: Non rotating, rotating, non rotating with late mixing, and rotating with late mixing. Other chemical elements, such as Na, are discussed. The goal is to find a distinct nucleosynthetic signature for each category of model. It should allow to attribute a preferred source star to an observed CEMP-no star and then further test the possible scenarios (especially the late mixing scenario).

We mainly study [X/H] ratios (where X is C, N, O, or Na etc.) and then use them to go further in the comparison between models and observations (c.f. Sect. \ref{sec:8}). [X/H] ratios likely give the absolute amount of elements since the abundance variation of H is modest.

\subsubsection*{\textit{Non rotating models without late mixing}}

The outer layers of the non-rotating models without late mixing have a [X/H] pattern generally close to the initial ISM pattern. They do not synthesise primary $^{13}$C and $^{14}$N. A low [N/H] is then a signature of this class of models. Also a high $^{12}$C/$^{13}$C ratio might be a signature, although this ratio can nevertheless be low in the layers of the stars where the CNO cycle has operated (see Fig. \ref{CNCC_norot}). Globally, however, a high $^{12}$C/$^{13}$C is favoured since no primary $^{13}$C is synthesised.

\subsubsection*{\textit{Rotating models without late mixing}}

Because of the rotational mixing that allows some exchanges of material between the H- and He-burning regions, rotating models synthesise a variety of isotopes \citep[see e.g.][]{maeder15b, choplin16pap}. Among them $^{13}$C and $^{14}$N are created because $^{12}$C and $^{16}$O diffuse from the He core to the H shell. Some primary $^{14}$N is then engulfed by the He-burning region. The chain $^{14}$N($\alpha,\gamma$)$^{18}$F($e^+ \nu_\text{e}$)$^{18}$O($\alpha,\gamma$)$^{22}$Ne boosts the $^{22}$Ne abundance in the He core compared to the non rotating case. When rotation is sufficiently fast, some $^{22}$Ne can diffuse back to the H-burning shell so that the Ne-Na cycle is boosted and some extra $^{23}$Na is synthesised. We call it the \textit{first channel} of $^{23}$Na production. To a lesser extent, the Mg-Al cycle is also enhanced. Close to the end of the core He-burning phase, some $^{22}$Ne in the He-burning core is transformed into $^{25}$Mg and $^{26}$Mg through $^{22}$Ne($\alpha,n$)$^{25}$Mg and $^{22}$Ne($\alpha,\gamma$)$^{26}$Mg. At this point, only a short time is left until the end of the evolution but some Mg can still diffuse in the H shell, boosting the Mg-Al cycle so that more Al can be created. The neutrons released by $^{22}$Ne($\alpha,n$)$^{25}$Mg can be captured by $^{22}$Ne to form $^{23}$Na through $^{22}$Ne($n,\gamma$)$^{23}$Ne($\beta^{-}\bar{\nu}_\text{e}$)$^{23}$Na. We call this the \textit{second channel} of $^{23}$Na production. The $^{23}$Na synthesised in this way cannot be ejected without being accompanied by He-burning products, which would drastically increase the $^{12}$C/$^{13}$C ratio, in contrast with what is observed. Also the s-process is boosted in the He core compared to non-rotating models. Indeed, more neutrons are released since the abundance of $^{22}$Ne is higher. The s-elements could be an interesting complementary nucleosynthetic signature, likely differing between the categories of models. Although the present work focuses on lighter elements, we plan to investigate the s-process in the frame of the CEMP stars in a future work. To summarise, the main signatures of this category of model would be a relatively high [N/H] and [Na/H]. In addition, a lower $^{12}$C/$^{13}$C is expected compared to non-rotating models because of the synthesis of primary $^{13}$C. 

\subsubsection*{\textit{Non-rotating models with late mixing}}

These models synthesise a lot of primary $^{13}$C and $^{14}$N. The late mixing boosts the CNO cycle so that C, N, and O are strongly processed in the H shell but the equilibrium is not necessarily reached. The amount of primary $^{13}$C and $^{14}$N produced by the late mixing event is higher than the amount produced with pure rotational mixing. The other elements (especially Na, see explanation in the previous and following paragraphs) remain at lower abundances than in rotating models. The characteristic [X/H] pattern for these models would be a high CNO enhancement but with few other light elements and a higher [C/N] ratio than non-rotating and rotating models.

\subsubsection*{\textit{Rotating models with late mixing}}

As in the previous category, the CNO cycle is boosted a lot in these models. Because of the late mixing, some $^{23}$Na, synthesised through the \textit{second channel} (c.f. previous discussion about rotating models), is transferred from the He-burning to the H-burning shell, so [Na/H] is enhanced in the H shell. $^{12}$C/$^{13}$C is low because of the late mixing that synthesises a lot of primary $^{13}$C (c.f. the previous category). To summarise, these models would present the same signature as the previous category but with a lot more Na.\\

An interesting point is that a high Na abundance can only be achieved with a progressive mixing at work during the whole evolution of the source star. This is because the source of $^{23}$Na in both the H (\textit{first channel}) and He shell (\textit{second channel}) is $^{22}$Ne, which is largely boosted if the progressive mixing operates. A natural candidate for the progressive mixing is the rotational mixing. 

\section{Connecting the CEMP-no with their source star(s)}
\label{sec:7}

\begin{table*}
\scriptsize{
\caption{CEMP-no stars with $-3.5 <$ [Fe/H] $< -4.1$, [C/Fe] $>0.7$ and [Ba/Fe] $<1$. \citep[Taken mainly from the SAGA database,][]{suda08}. The stars are classified as MS if $T_\text{eff} > 5500$ K and $\log g \geq 3.25$ \label{table:2} and as RGB otherwise.} 
\begin{center}
\resizebox{18.5cm}{!} {
\begin{threeparttable}
\begin{tabular}{llrrrrrrrrrrl}
\hline
\hline
 Star               & Type & [Fe/H] & A(Li)  & [C/H]  & [N/H]  & [O/H]  & [Na/H] & [Mg/H] & [Al/H] & [Si/H] & $^{12}$C/$^{13}$C & Ref         \\
 \hline
 CS29498-043        & RGB  & -3.85  & \ensuremath{<}-0.05 & -1.13  & -2.14  & -1.48  & -2.82  & -2.07  & -3.1   & -2.77  & 8.0      & 1,2         \\
 CS29527-015        & MS   & -3.55  & 2.07   & -2.37  & -      & -      & -3.75  & -3.12  & -3.5   & -3.4   & -        & 3,4,5,6,7,8 \\
 G77-61             & RGB  & -4.0   & \ensuremath{<}1.16  & -0.8   & -1.8   & -2.2   & -3.4   & -3.51  & -      & -      & 5.0      & 9,10        \\
 HE0134-1519        & RGB  & -3.98  & 1.27   & -2.98  & \ensuremath{<}-2.98 & \ensuremath{<}-1.08 & -4.22  & -3.73  & -4.36  & -3.93  & \ensuremath{>}4.0     & 11          \\
 HE1012-1540        & RGB  & -3.76  & \ensuremath{<}0.75  & -1.77  & -3.02  & \ensuremath{<}-1.76 & -2.11  & -1.96  & -3.07  & -3.2   & \ensuremath{>}30.0    & 1,2,12      \\
 HE2331-7155        & RGB  & -3.68  & \ensuremath{<}0.37  & -2.34  & -1.11  & \ensuremath{<}-1.98 & -3.22  & -2.48  & -4.06  & -      & 5.0      & 11          \\
 HE0049-3948        & RGB  & -3.68  & -      & \ensuremath{<}-1.87 & \ensuremath{<}-1.28 & -      & -3.76  & -3.39  & -      & -3.69  & -        & 13          \\
 HE0057-5959        & RGB  & -4.08  & -      & -3.22  & -1.93  & -      & -2.1   & -3.57  & -      & -      & \ensuremath{>}2.0     & 13,14       \\
 HE0228-4047        & MS   & -3.75  & -      & \ensuremath{<}-1.87 & -      & -      & -      & -3.49  & -      & -3.4   & -        & 13          \\
 HE0945-1435        & MS   & -3.78  & -      & \ensuremath{<}-2.08 & -      & -      & -      & -3.88  & -      & \ensuremath{<}-1.78 & -        & 13          \\
 HE1201-1512        & MS   & -3.92  & -      & -2.78  & \ensuremath{<}-2.69 & -      & -4.27  & -3.72  & -4.65  & -      & \ensuremath{>}20.0    & 13,14       \\
 HE1346-0427        & MS   & -3.58  & -      & \ensuremath{<}-2.48 & -      & -      & -3.73  & -3.33  & -3.65  & \ensuremath{<}-1.78 & -        & 12,13       \\
 HE1506-0113        & RGB  & -3.54  & -      & -2.07  & -2.93  & -      & -1.89  & -2.65  & -4.07  & -3.04  & \ensuremath{>}20.0    & 13,14       \\
 HE2032-5633        & MS   & -3.63  & -      & \ensuremath{<}-1.27 & \ensuremath{<}-1.03 & -      & -3.72  & -3.34  & -4.14  & -3.3   & -        & 13          \\
 HE2139-5432        & RGB  & -4.02  & -      & -1.43  & -1.94  & -      & -1.87  & -2.41  & -3.66  & -3.02  & \ensuremath{>}15.0    & 13,14       \\
 HE2318-1621        & RGB  & -3.67  & -      & -2.63  & -2.43  & -      & -2.96  & -3.47  & -4.25  & -      & -        & 15          \\
 SDSSJ161956+170539 & MS   & -3.57  & -      & -1.35  & -      & -      & -      & -3.53  & -      & -3.88  & -        & 16          \\
 SDSSJ2209-0028     & MS   & -3.96  & -      & -1.4   & -      & -      & -      & -      & -      & -      & -        & 17          \\
 53327-2044-515\tnote{a}    & $-$  & -4.05  & -      & -2.7   & -      & \ensuremath{<}-1.24 & -3.91  & -3.65  & -4.22  & -      & \ensuremath{>}2.0     & 13,14          \\
 Segue1-7           & RGB  & -3.52  & -      & -1.22  & -2.77  & \ensuremath{<}-1.31 & -2.99  & -2.58  & -3.29  & -2.72  & \ensuremath{>}50.0    & 14          \\
\hline
\end{tabular}
\begin{tablenotes}
            \item[a] The evolutionary status of 53327-2044-515 is uncertain so that the average abundances of dwarf and subgiant solutions are taken, as done in \cite{norris13}.
\end{tablenotes}
\end{threeparttable}
}

\end{center}

\textbf{References}. 1 - \cite{roederer14a}; 2 - \cite{roederer14c}; 3 - \cite{spite12}; 4 - \cite{bonifacio07}; 5 - \cite{bonifacio09}; 6 - \cite{andrievsky07}; 7 - \cite{andrievsky10}; 8 - \cite{andrievsky08}; 9 - \cite{beers07}; 10 - \cite{plez05}; 11 - \cite{hansen15}; 12 - \cite{cohen13}; 13 - \cite{yong13}; 14 - \cite{norris13}; 15 - \cite{placco14a}; 16 - \cite{caffau13}; 17 - \cite{spite13}

}
\end{table*}

Here we discuss the case of some CEMP-no stars and make an attempt to select their most likely source star(s). We investigate 20 stars having [C/Fe] $> 0.7$, [Ba/Fe] $< 1$, and [Fe/H] $= - 3.8 \pm 0.3$ (see Table \ref{table:2}). We have selected the CEMP-no stars having a narrow range of [Fe/H] so that the source-star models can be computed with only one metallicity. Since the observed $^{12}$C/$^{13}$C ratios imply that only the outer regions of the source-star models should be expelled (the inner regions, and thus all the iron produced, being locked into the remnant), we use the same [Fe/H] for the source stars as for the CEMP stars. 

\subsection{Correcting for the effect of the first dredge up in CEMP-no stars}

To compare the predicted and observed [X/H] ratios, we need to know the initial [X/H] surface ratios of the CEMP-no stars, the ones that likely reflect the abundances of their natal cloud, hence the ratios in the material ejected by the source star (We note that the dilution of the source star ejecta with the ISM can be important; it is discussed in Sect.~\ref{sec:dilution}). Non-evolved CEMP-no stars have a surface composition that is probably very close to the initial one. In more-evolved stars, the first dredge up can have occurred, likely reducing the surface carbon abundance. When available for the considered CEMP-no star, we have taken into account the correction on [C/Fe] predicted by \cite{placco14c}. For 505 metal-poor stars, they have determined the correction to apply to this ratio in order to recover the initial ratio. This correction, generally lower that 1 dex, remains small compared to the dispersion of observed [C/Fe].

\subsection{The atomic diffusion}

In addition to the dredge up, the atomic diffusion adds another source of uncertainty when one wishes to link a CEMP-no with its possible source star. The effect of atomic diffusion (gravitational settling, thermal diffusion, and radiative acceleration) in low-mass metal-poor stars has been studied by \cite{richard02}, for example. They predict changes ranging between $\sim 0.1$ and $1$ dex, depending on the chemical species considered, $T_\text{eff}$ and on the evolutionary status of the model (c.f. their figure 13 and 14). \cite{richer00} have shown that an additional turbulence seems to be required to explain the chemical anomalies of AmFm stars. When such an additional turbulence is considered in the metal-poor models of \cite{richard02}, the effect of the atomic diffusion on the surface abundances is significantly reduced; around $\sim 0.1$ dex in many cases, when the turbulence is strong enough. The effect of atomic diffusion on the surface abundances of CEMP-no stars cannot be very well predicted but there are hints suggesting that it is modest. 

\subsection{Dilution of the ejecta with the ISM\label{sec:dilution}}

A third aspect to take into account before comparing the source-star ejecta with the CEMP stars is that the ejecta could have been diluted in the ISM. In that case, the CEMP-no star would be made of a mixture of ejecta plus initial ISM\footnote{By initial ISM we mean here the ISM in which the source star formed.}. We need to ascertain the degree of dilution. As discussed in \cite{meynet10}, we use the Li abundance for this purpose: Li is a fragile element, so it is completely destroyed in the ejecta of the source star. When Li is observed at the surface of the CEMP-no stars, it must come from the dilution with the initial ISM. Here we suppose that the Li abundance in the initial ISM is equal to the WMAP value of 2.72 \citep{cyburt08}. Knowing how much mass is ejected from the source star, one can find the mass of initial ISM $M_\text{ISM}$ to add to the ejected mass $M_\text{ej}$ in order to reproduce the Li abundance observed at the surface of the CEMP-no star. The dilution factor $D$ is then  expressed as
\begin{equation}
D = \frac{M_\text{ISM}}{M_\text{ej}} =  \frac{X(Li)_\text{CEMP}}{X(Li)_\text{ISM} - X(Li)_\text{CEMP}}
,\end{equation}
where $X(Li)_\text{CEMP}$ and $X(Li)_\text{ISM}$ are the mass fraction of Li at the surface of the CEMP-no star and in the initial ISM, respectively. A simple example is if the CEMP-no star has no Li. In this case, $D=0$ and it would be made of pure source star ejecta. If the Li abundance is instead very close to the WMAP value, much more initial ISM would be needed to form this CEMP-no star. When the considered CEMP-no star has a measured Li abundance, we apply this method and dilute the source-star ejecta with the corresponding mass of initial ISM. If no Li data is available, we assume $D=0$ by default. We also discuss the impact of having $D>0$ in some cases, even without Li measurement. We do not pretend to give the right value of $D$ for the CEMP-no stars but rather try to discuss the impact of dilution.

We have assumed that the surface Li abundance has not changed since the birth of the CEMP-no star. If in situ processes have occurred and destroyed some Li, it would mean that $X(Li)_\text{CEMP}$ should be corrected to recover the initial Li content. This point is discussed for the star HE2331-7155.

   \begin{figure*}[t]
   \centering
   \begin{minipage}[c]{.49\linewidth}
       \includegraphics[scale=0.46]{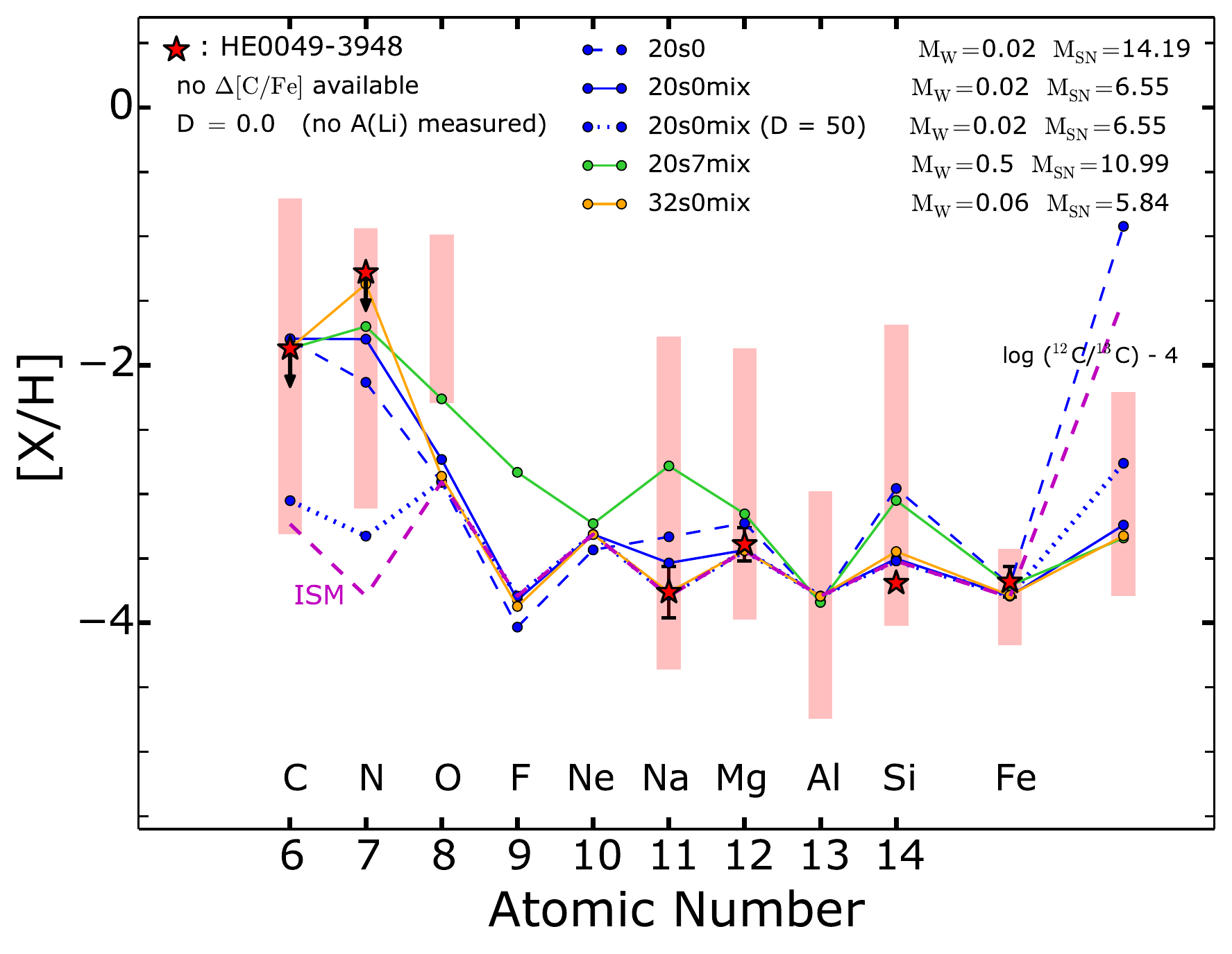}
   \end{minipage}
   \begin{minipage}[c]{.49\linewidth}
      \includegraphics[scale=0.46]{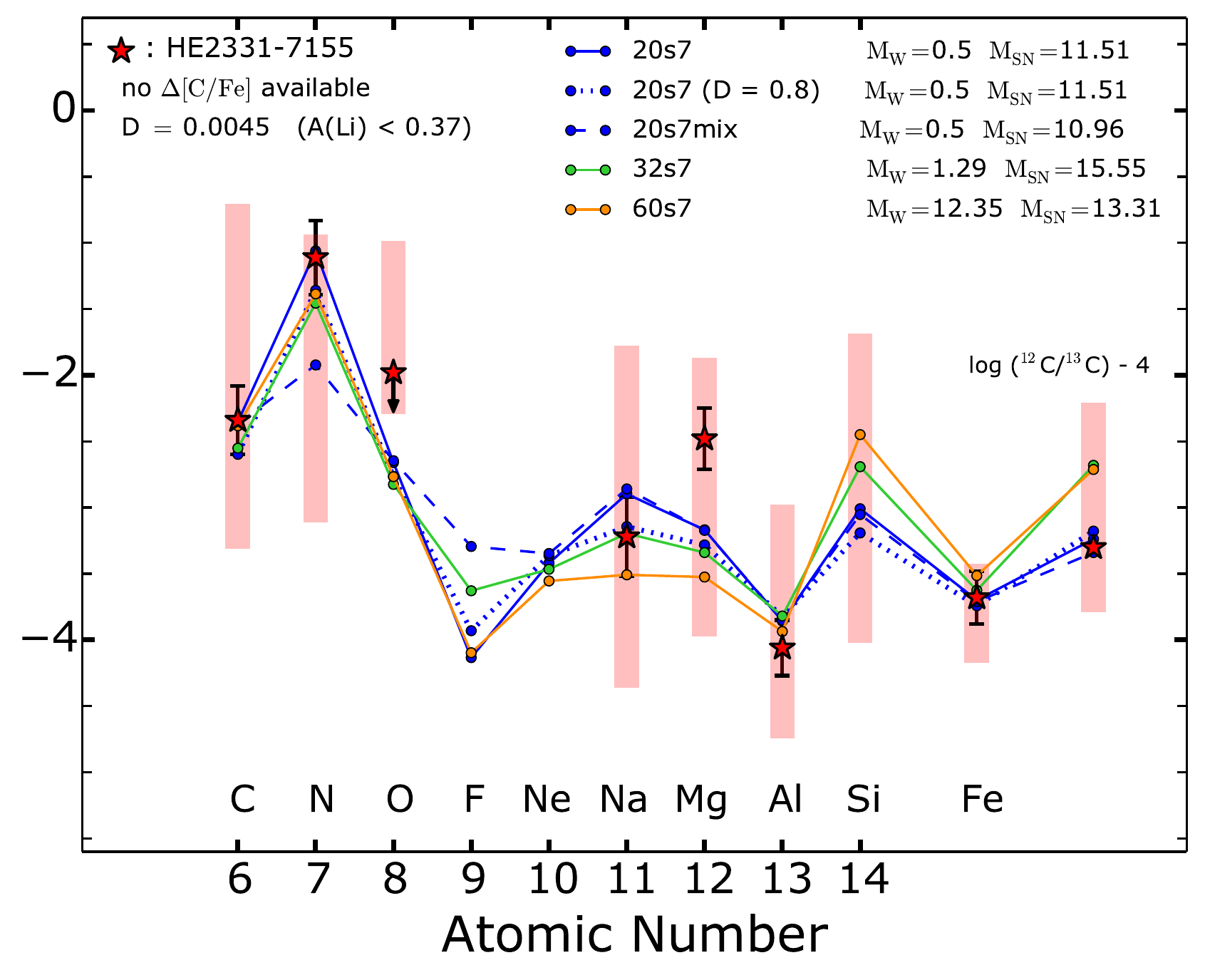}
   \end{minipage}
   \begin{minipage}[c]{.49\linewidth}
      \includegraphics[scale=0.46]{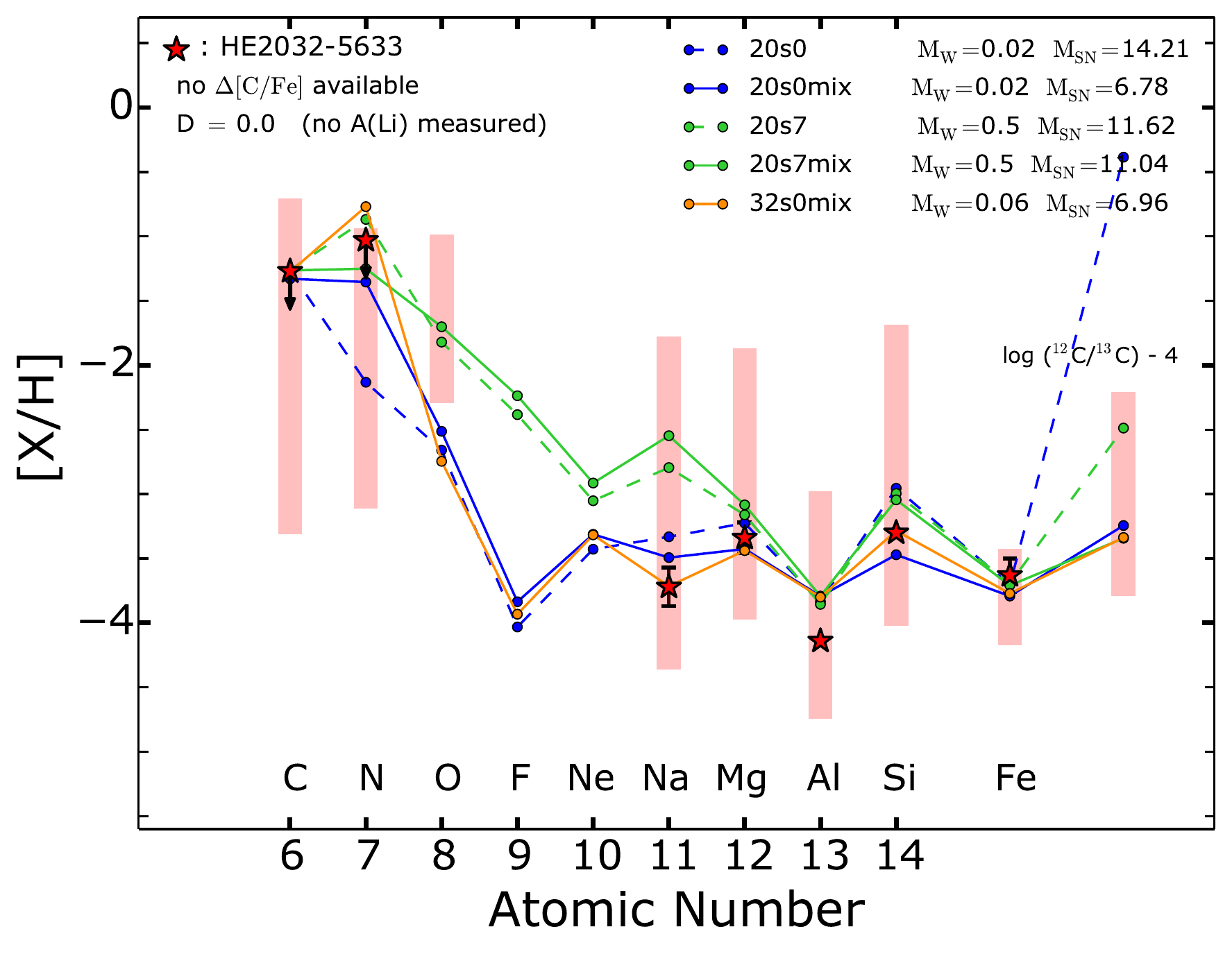}
   \end{minipage}
   \begin{minipage}[c]{.49\linewidth}
      \includegraphics[scale=0.46]{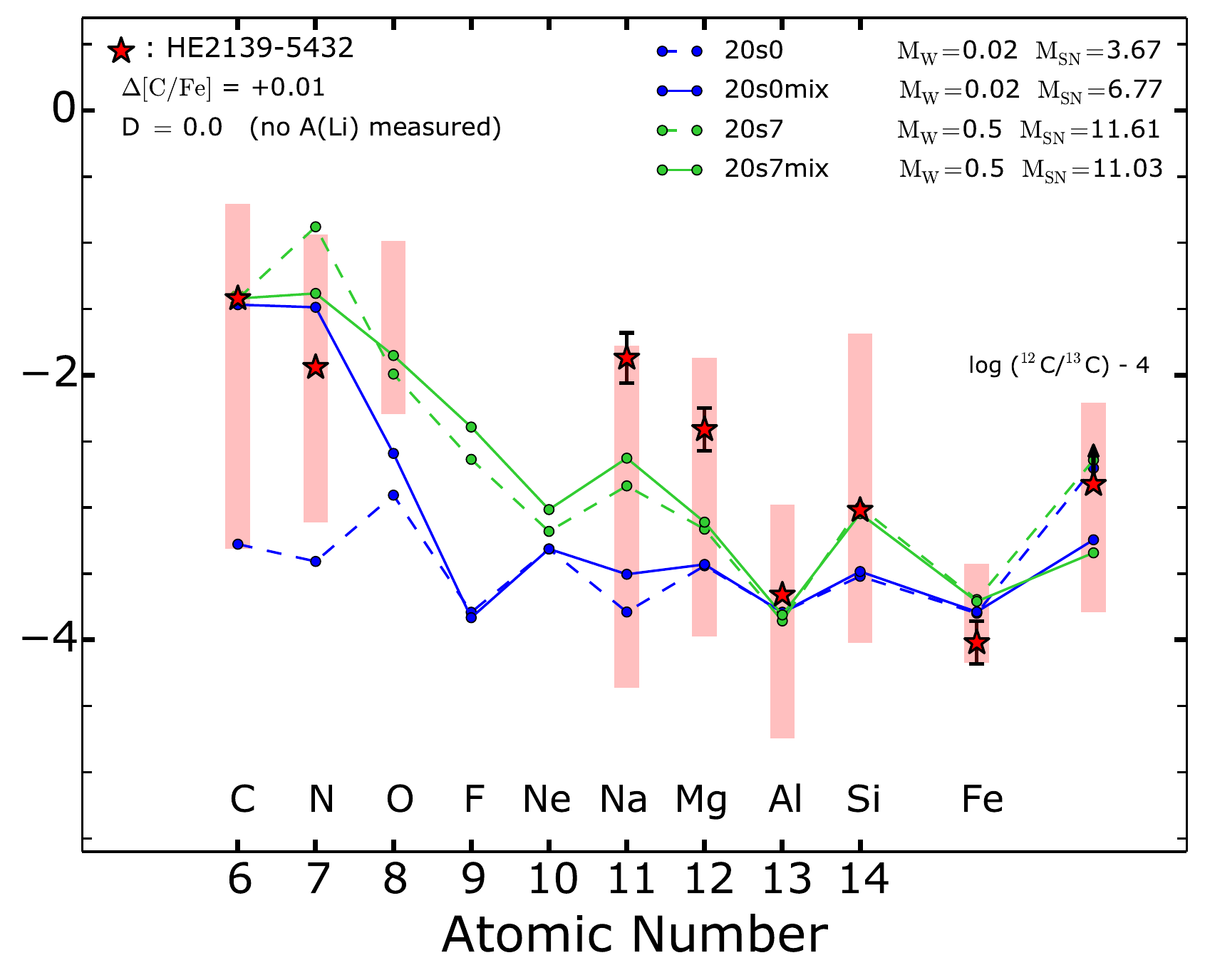}
   \end{minipage}
   \begin{minipage}[c]{.49\linewidth}
      \includegraphics[scale=0.46]{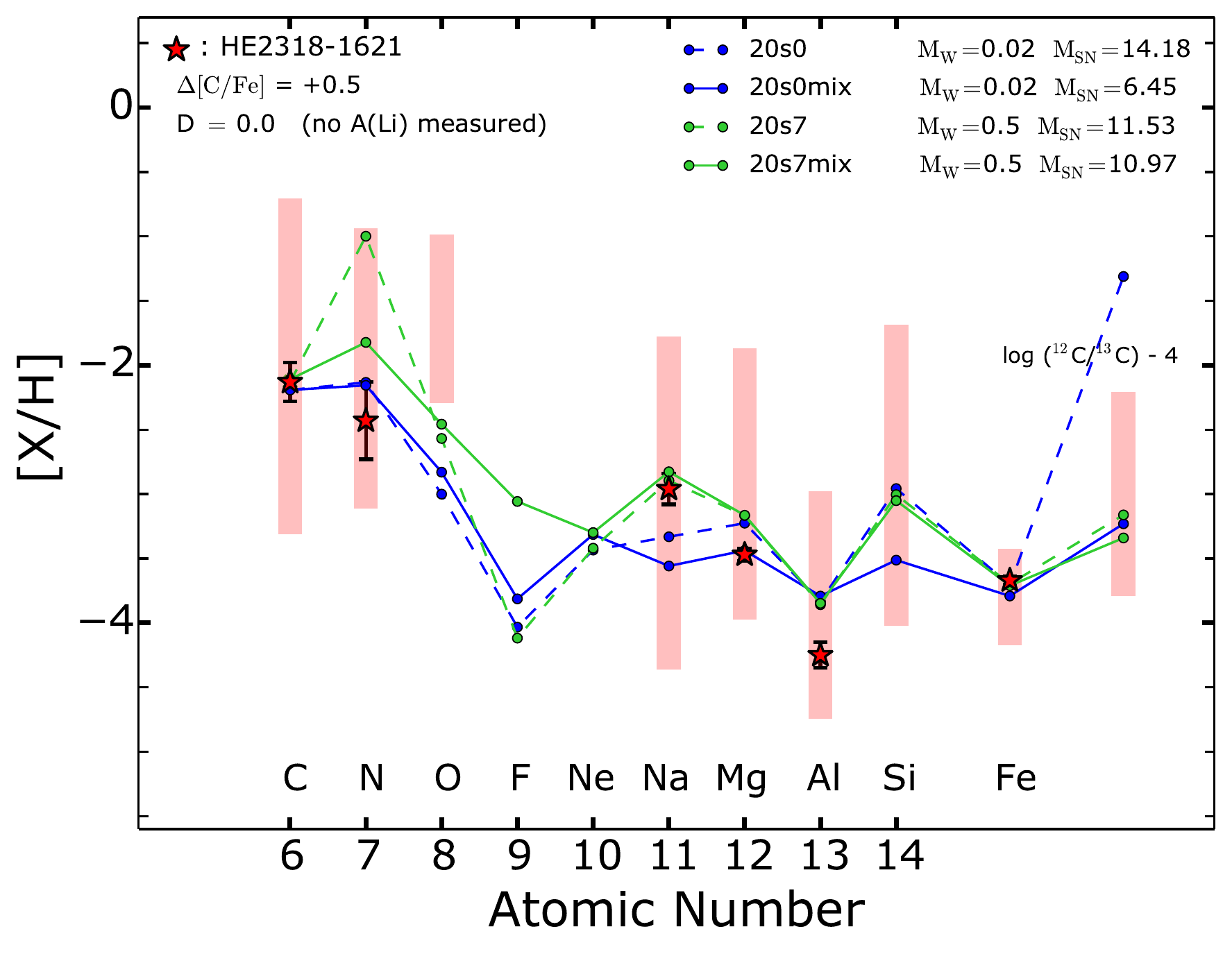}
   \end{minipage}
   \begin{minipage}[c]{.49\linewidth}
      \includegraphics[scale=0.46]{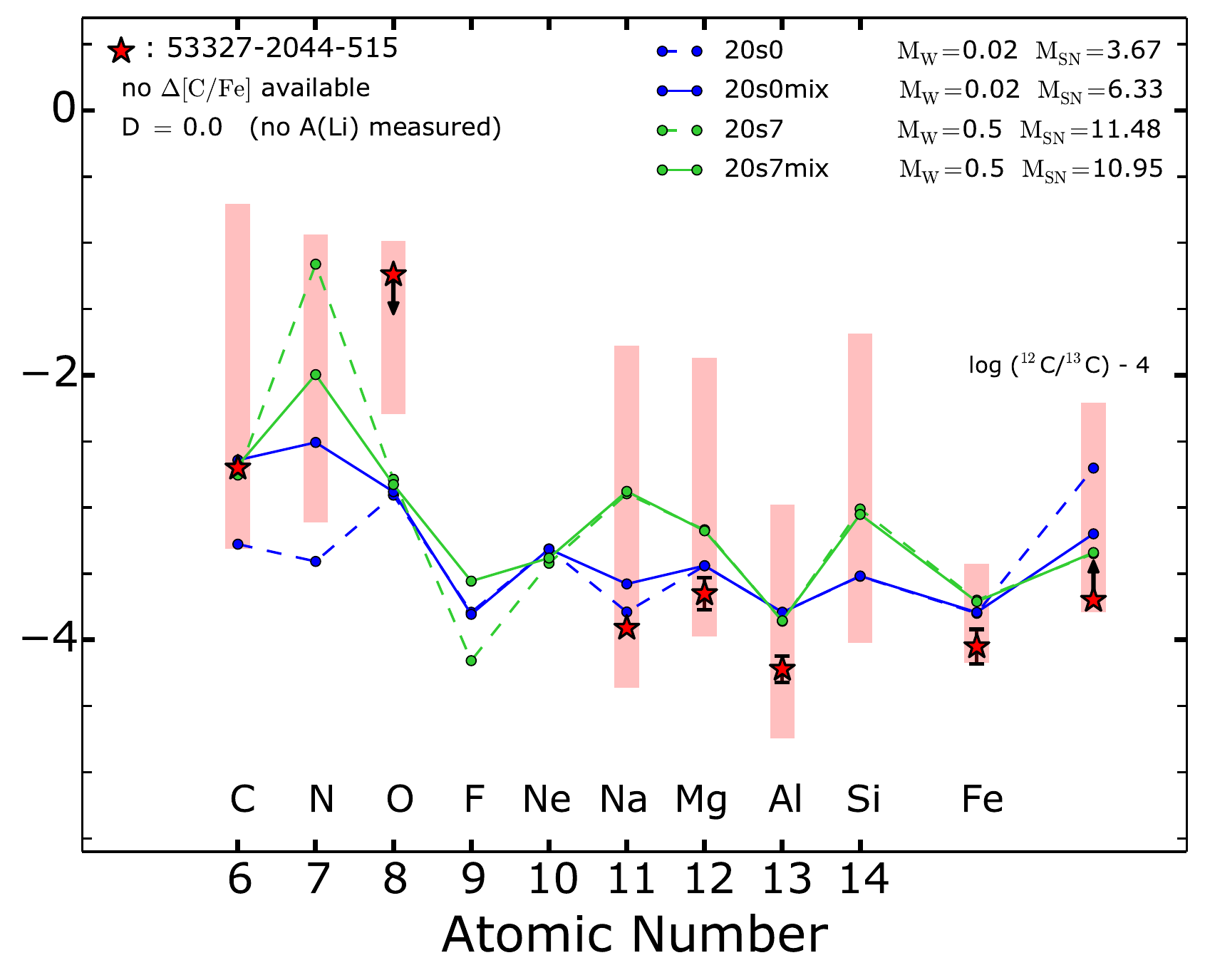}
   \end{minipage}
   \caption{[X/H] as a function of the atomic number. Also log $^{12}$C/$^{13}$C is shown (shifted downwards by 4 dex for clarity). Each panel is dedicated to one observed CEMP-no star (red starry symbols). Arrows indicate upper/lower limits. If available, error bars are indicated. The correction $\Delta$[C/Fe] of \cite{placco14c} applied on [C/Fe] (if any) is taken into account and indicated in the upper-left corner. In each panel we show the composition of the ejecta of the indicated models (solid, dashed and dotted lines). The masses ejected by the winds ($M_\text{W}$) and the supernova ($M_\text{SN}$) are written in the upper-right corner of each panel. The composition of the ISM is shown in the first panel. The dilution coefficient $D$ and A(Li) value (if available) are shown in the upper-left corner. The range of observed [X/H] ratios and $\log(^{12}\text{C}/^{13}\text{C})$ ratios for the stars in Table \ref{table:2} are shown by the red rectangles.}
\label{star1}
    \end{figure*}

\subsection{Condition on the mass cut, weak explosion}

The source-star ejecta is composed of the wind ejected during the evolution plus a supernova with a $M_\text{cut}$ defined such that
\begin{equation}
\label{crit}
| \log(\rm ^{12}C/^{13}C)_\text{obs} - \log(^{12}C/^{13}C)_\text{mod}  | + | [C/H]_\text{obs} - [C/H]_\text{mod}  |
\end{equation}
is minimal. In this expression, `obs' refers to the observed value and `mod' to the value coming from the models. $^{12}$C/$^{13}$C generally gives tight constraints on the mass cut because this ratio varies considerably between the different burning zones of the source star; it goes from $\sim 5$ to $\sim \infty$ when going from the H- to the He-burning shell. The [C/H] ratio is also taken into account in the above criterion because sometimes there is no $^{12}$C/$^{13}$C value available and also because relying on $^{12}$C/$^{13}$C only can give multiple mass cut possibilities. This parametrisation and its consequences are discussed in Sect. \ref{faint}.

We note that there is no degeneracy between the effect of the mass cut and the dilution. Expelling deep layers, where He has burnt, raises [C/H]. Increasing $D$ allows one to reduce [C/H]. However, $^{12}$C/$^{13}$C increases both when deep layers are expelled and when $D$ increases meaning that a solution implying deep layers with significant dilution will always lead to a high $^{12}$C/$^{13}$C, likely out of the observed range.

\subsection{Comparison of the source-star models with observed CEMP-no stars}

We now more closely  inspect six CEMP-no stars, the others being discussed together in Sect. \ref{others}. Each of the six panels of Fig. \ref{star1} is dedicated to one star (represented by the red starry shaped symbols) together with the composition of the source-star model ejecta.

We focus on light elements: C, N, O, Na, Mg, Al and Si. By definition, the CEMP-no stars are generally not enhanced in heavier elements (especially s- and r-elements). As we  see below, in some stars, heavy elements, such as Sr or Ba for instance, are present, and although they are only in small amounts, they should be explained (c.f. discussion about 53327-2044-515).

\subsubsection*{HE0049-3948}

This star (see the upper-left panel of Fig. \ref{star1}) has only upper limits available for C and N so some caution is required regarding these elements. The low sodium abundance favours non-rotating models; [Na/H] is indeed too high when considering the 20s7mix model (green pattern). The silicon abundance is also low. This can be achieved if only the outer layers of the source star are expelled, because in deeper layers, the temperature is higher and more silicon is synthesised through the Mg-Al-Si chain. We see, indeed, that the models expelling the largest amount of mass (see $M_\text{W}$ and $M_\text{SN}$ in Fig. \ref{star1}), that is, the 20s0 (blue dashed line) and 20s7mix (green line) models, have an overly high [Si/H]. Constraining the mass of the source star could be done through a comparison between the predicted and observed C and N abundance and $^{12}$C/$^{13}$C ratio (c.f. Fig. \ref{CNCCmix} for instance). The determination of these abundances at the surface of this star would be required. We note that our models reproduce the upper limit of the [C/H] ratio. Of course lower values can be obtained by varying the mass cut for instance.

The heavier elements observed in this star are little or not enhanced. For instance, [Sr/Fe] $= - 0.85$ and [Ba/Fe] $<0.14$ \citep{yong13}.

No  Li data is available so $D$ cannot be constrained. By default we have taken $D=0$. The case $D=50$ (it would imply A(Li) $=2.71$) for the ejecta of the 20s0mix model is shown by the dotted blue line; it is shifted towards the ISM pattern (magenta line). The case of a large dilution factor cannot be discarded because of the upper limits on C and N and because although raised, $^{12}$C/$^{13}$C remains within the observed range.

\begin{table*}
\scriptsize{
\caption{Preferred source star(s) of the six CEMP-no stars discussed above. 'yes' indindicates that the source star is the (or among the) preferred one(s), 'yes/no' indicates the possible candidates for which caution is needed or some additional observational data would be needed to reach stronger conclusions. \label{table:3}} 
\begin{center}
\resizebox{18.3cm}{!} {
\begin{tabular}{l|ccc|ccc|ccc|ccc} 
\hline % inserts double horizontal lines
\hline % inserts single horizontal line
 &  \multicolumn{3}{c|}{\textbf{No Rot. models}} &  \multicolumn{3}{c|}{\textbf{Rot. models}}  & \multicolumn{3}{c|}{\textbf{No Rot. mix models}} & \multicolumn{3}{c}{\textbf{Rot. mix models}}\\ % table heading
Star  & 20s0 & 32s0 & 60s0  & 20s7 & 32s7 & 60s7 & 20s0mix & 32s0mix & 60s0mix& 20s7mix & 32s7mix & 60s7mix   \\
\hline % inserts single horizontal line
\textbf{HE0049-3948}    &  no     &     no      &       no      &  no & no      &       no      &       \textbf{yes}    &       \textbf{yes}    &       \textbf{yes}    &       no      &  no             &       no      \\
\textbf{HE2331-7155}    &  no     &     no      &       no      &  \textbf{yes} &       no      &       no      &       no      &       no      &       no              &       no      &  no             &       no      \\
\textbf{HE2032-5633}    &  \textbf{yes/no}        &     \textbf{yes/no} &       \textbf{yes/no} &  \textbf{yes/no} &      \textbf{yes/no} &       \textbf{yes/no} &       \textbf{yes}    &       \textbf{yes/no} &       \textbf{yes/no} &       \textbf{yes/no} &  \textbf{yes/no}                &       \textbf{yes/no} \\
\textbf{HE2139-5432}    &  no     &     no      &       no      &  no & no      &       no      &       no      &       no      &       no              &       \textbf{yes/no} &  no             &       no      \\
\textbf{HE2318-1621}    &  no     &     no      &       no      &  no & no      &       no      &       \textbf{yes     }       &       no      &       no              &       \textbf{yes}    &  no             &       no      \\
\textbf{53327-2044-515}&  no &          no      &       no      &  no & no      &       no      &       \textbf{yes/no} &       \textbf{yes/no} &       \textbf{yes/no} &       no      &  no             &       no      \\
\hline

\end{tabular}
}
\end{center}
}

\end{table*}

\subsubsection*{HE2331-7155}

This star (upper-right panel in Fig. \ref{star1}) has a high [N/H] and [C/N] $= -1.23$. The high [N/H] clearly shows the need for a mixing event (progressive or brutal) between the H and He-burning regions of the source star. The low [C/N] discards the models with late mixing. At the same time, the low $^{12}$C/$^{13}$C favours models with mixing (rotational or late), where primary $^{13}$C is formed. These points, together with the relatively high Na and high Mg point towards models with a progressive mixing, achieved here by rotation. The best source star for HE2331-7155 is the 20s7 model (represented by the solid blue curve). This model provides enough N together with a low $^{12}$C/$^{13}$C. The 32s7 and 60s7 models (green and orange solid lines) do not provide enough primary N and have a higher $^{12}$C/$^{13}$C ratio. The Na, Al and $^{12}$C/$^{13}$C are also rather well explained by the 20s7 model. Only the Mg is underproduced by $\sim 1$ dex. However, the nuclear rates implying Mg at relevant temperature in the H-burning shell (30 - 80 MK) are not very well known. Changing these rates can lead to significant differences in the nucleosynthesis and thus in the composition of the ejecta \citep[see e.g.][]{decressin07,choplin16pap}. Future laboratory measurements of these rates will allow a more accurate comparison of the observed Mg/H value with the value predicted by stellar models. 
A(Li) $< 0.37$ for that star. Taking 0.37 for A(Li) gives $D = 0.0045$, which is small and then barely changes the composition of the source-star ejecta. Let us suppose that this CEMP-no star has destroyed 2 dex of Li at its surface since its birth\footnote{The models of \cite{korn09} predict a maximum Li depletion of 1.2 dex for the star HE1327-2326. Here 2 dex would likely correspond to an extreme case.} so that its initial A(Li) would be 2.37. In this case, $D =0.8$. The 20s7 model with such a dilution factor is shown by the blue dotted pattern. Whatever the species considered, it implies a shift of less than 0.5 dex compared to the $D=0.0045$ case. This is because the initial ISM is much more metal poor than the material ejected from the source star; it shows the small effect of the dilution even if the initial Li content in the CEMP-no $-$ hence $D$ $-$ was higher.

The heavier elements observed on this star are little or not enhanced. Especially, [Sr/Fe] $= - 0.85$ and [Ba/Fe] $=-0.90$ \citep{hansen15}.  

\subsubsection*{HE2032-5633}

The upper limits on C and N together with the absence of $^{12}$C/$^{13}$C measurement prevent any strong conclusion regarding this star (middle-left panel of Fig. \ref{star1}). The $^{12}$C/$^{13}$C ratio of the 20s0 model is $\sim 2$ dex above the observed range. The other standard non-rotating models behave in a similar way. However, as long as no $^{12}$C/$^{13}$C ratio is observed at the surface of this star, we cannot discard these models.

Also, as we see by comparing the blue and green patterns, the low [Na/H] ratio on HE2032-5633 would favour non-rotating models. Finally, the 20s0mix model is the best source star candidate for this CEMP-no star. The 32s0mix has an overly high [N/H] ratio (see the orange pattern). However, we remain very cautious, since the mass cuts of the models are set to reproduce the observed [C/H]. Here, [C/H] is only an upper limit. A lower measured [C/H] ratio would imply a smaller amount of ejected mass (smaller $M_{SN}$) and would change the abundances in the material ejected by the source star models.

There are also few heavy elements at the surface of HE 2032-5633: [Sr/Fe] $<-0.68$ and [Ba/Fe] $<0.31$ \citep{yong13}.

\subsubsection*{HE2139-5432}

This star (middle-right panel of Fig. \ref{star1}) has [C/N] $= 0.51$ meaning that the models with late mixing are favoured. The 20s7mix model has the highest [C/N] (see Fig. \ref{CNCCmix}), closer to the observed [C/N] than the 32s7mix and 60s7mix models. [Na/H] is the highest in the sample and [Mg/H] is high as well. This cannot be explained at all by non-rotating models. The high Na, Mg, and [C/N] clearly point towards the rotating models with late mixing. The 20s7mix model provides an interesting solution. We note however two discrepancies: (1) Na and Mg are underestimated by about $\sim$ 1 dex and (2) [N/H] is $\sim 0.5$ dex too high, while $^{12}$C/$^{13}$C is too low by at least $\sim 0.5$ dex. A solution to the first discrepancy could be to consider a faster rotator, since more mixing would enhance Na and possibly Mg (c.f. Sect. \ref{sec:6}). A solution for the second discrepancy would be to have a material less processed by the CNO cycle in the partially processed zone. In this case, we would have less $^{14}$N and $^{12}$C/$^{13}$C might not have reached its CNO equilibrium value at the end of the evolution. This would be coherent with the [N/H] and $^{12}$C/$^{13}$C ratios of this CEMP-no star. Less CNO processing can be achieved in the source star if the late mixing event occurs closer to the end of evolution. Hence, to resolve (1) and (2) simultaneously, a possibility might be to consider a faster rotator that underwent a late mixing event very close to the end of its evolution. Even if the 20s7mix model does not perfectly match this CEMP-no star, rotational mixing with a late mixing process are likely to provide a solution.

This star is also poor in heavy elements: [Sr/Fe] $=-0.55$ and [Ba/Fe] $<-0.33$ \citep{yong13}.

\subsubsection*{HE2318-1621}

Because of its low $\log g$ and $T_\text{eff}$, \cite{placco14c} have proposed a correction $\Delta$[C/Fe] $= 0.5$ dex for this star (lower-left panel of Fig. \ref{star1}).
This implies [C/H] $=-2.13$ and [C/N] $=0.3$. The high [C/N] points towards models with late mixing. As for HE2139-5432, the 20s7mix model is preferred because of its high [C/N] in the partially CNO-processed zone, closer to the observed value than the 32s7mix and 60s7mix models. Also, the relatively high [Na/H] favours the rotating models. The 20s0mix model for instance provides a relatively good fit but underestimates [Na/H]. The 20s7mix provides the right amount of Na but slightly too much N. In any case, the $20\,M_\odot$ models with late mixing, both rotating and non-rotating, provide the best solutions. Since deeper layers are expelled from the rotating model, [Si/H] is higher in the ejecta by about 0.7 dex compared to the non-rotating model (compare the solid blue and green lines). Then, the silicon abundance of this star would be an interesting diagnostic to discriminate the two source stars and find the best one.

At the surface of this CEMP-no star, [Sr/Fe] $=-1$ and [Ba/Fe] $=-1.61$. Also, [Eu/Fe] $<0.13$ \citep{placco14a}.

\subsubsection*{53327-2044-515}

The low [Na/H] of this star disfavours the rotating models, as we see on the lower-right panel of Fig. \ref{star1}. Since the 20s0 model cannot provide both a high [C/H] and a low $^{12}$C/$^{13}$C, the 20s0mix model is preferred. We note however the lower limit for $^{12}$C/$^{13}$C. The O is largely underestimated (by about 2 dex) but since it is an upper limit, it is not incompatible with the low predicted [O/H].

At the surface of this star, [Sr/Fe] $=1.09$ or 0.53, depending on whether it is considered as a dwarf or a subgiant \citep{yong13, norris13}. Similarly, [Ba/Fe] $<0.34$ or $<0.04$. The modest enrichment in Sr (and eventually Ba) at the surface of this star might be explained by the weak s-process having operated in  the source star. Rotation boosts the weak s-process because of the extra $^{14}$N brought into the He core, transformed into $^{22}$Ne that afterwards releases neutrons \citep{frischknecht12,frischknecht16}. There is however a drawback: Rotating models are rather disfavoured because of the low observed [Na/H]. Source star models with an extended nucleosynthetic network are needed to make quantitative predictions. This will be done in the future.

\subsubsection*{General remarks}

Table \ref{table:3} summarises the above discussion on the six CEMP-no stars. We see that in most cases, the source stars with late mixing are preferred, even if there are some important uncertainties for some stars (see previous discussion). We found that four out of the six considered CEMP-no stars probably cannot be explained without the late mixing process in the source star. Two stars cannot be explained without a progressive mixing, that can be achieved by rotation. One star, HE2331-7155, probably cannot be explained through the late mixing. It is explained better by a $20\,M_\odot$ rapidly rotating source star with no late mixing. Globally, the lower-mass source stars ($20\,M_\odot$) are preferred compared to the higher mass ones ($60\,M_\odot$).

\subsection{Other CEMP-no stars}
\label{others}

\begin{figure}
   \centering
       \includegraphics[scale=0.485]{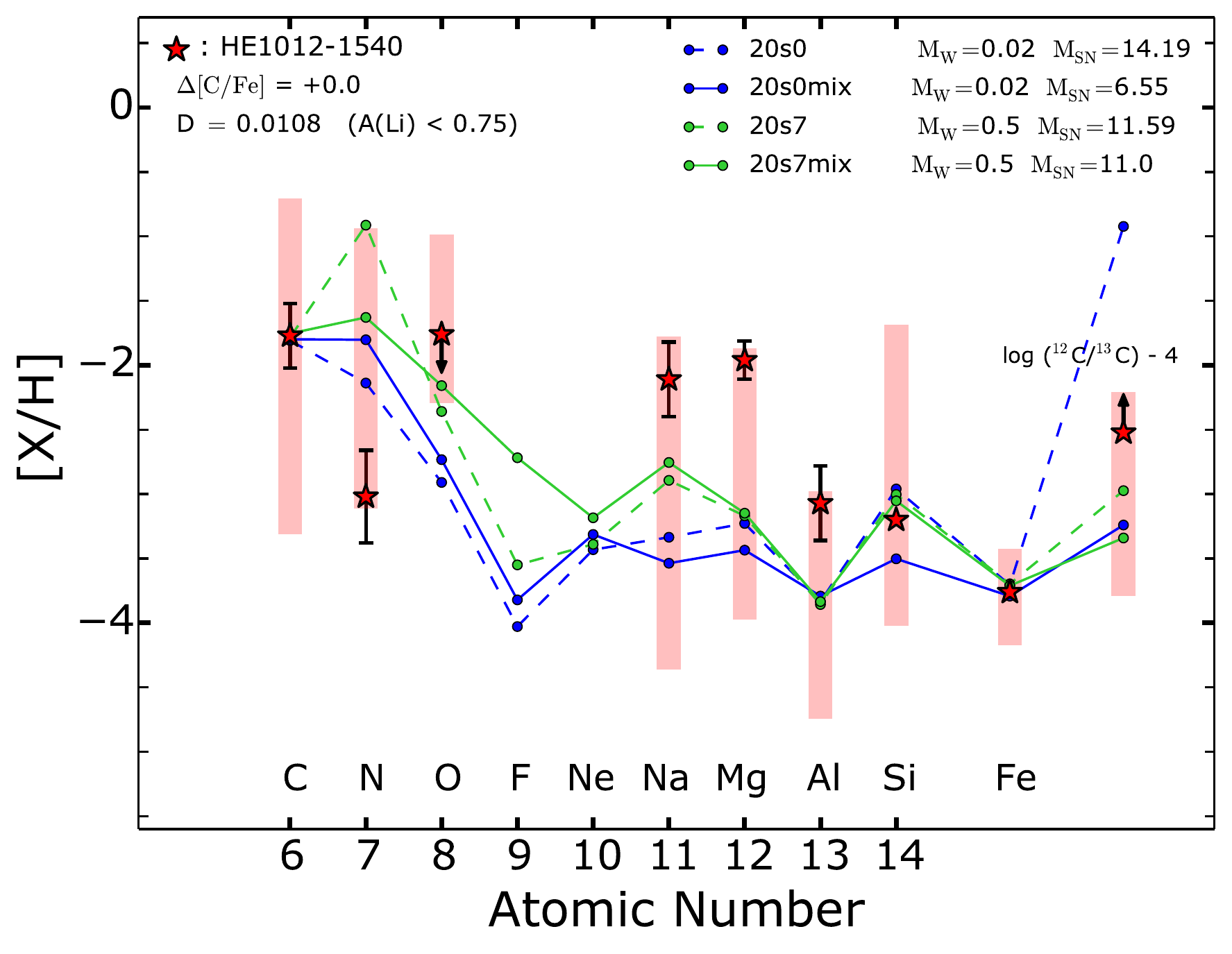}
   \caption{As in Fig. \ref{star1} but for HE1012-1540. The mass cut is calibrated to reproduce [C/H] only.}
\label{star7}
    \end{figure}

We have discussed in detail six out of the 20 stars in the sample (see Table \ref{table:2}). Here we discuss three points regarding the other stars:

\begin{enumerate}
\item Some of these stars have very little abundance data available (e.g. HE0945-1435, SDSSJ161956+170539 or SDSSJ2209-0028), meaning that different solutions could match the observations. More abundance data are needed to provide constraints and allow for a better estimation of the source star.

\item Several stars have very low [X/H] ratios, meaning that they might be formed with only a little amount of the source star ejecta and mainly with the ISM. For instance, HE0134-1519 and HE1201-1512 belong to the less enriched stars in the sample. HE0134-1519 is a RGB star with A(Li) $=1.27$. As discussed, the presence of Li might indicate an important contribution of the ISM. However, the initial Li content might be higher since this star is a RGB, in which internal processes could have depleted this element at the surface. In that case, the dilution factor would be higher. Also CS29527-015 has generally low [X/H] ratios. It is also the Li-richest star in the sample (A(Li) = 2.07), suggesting an important contribution of the ISM.

\item Finally, some stars are difficult to explain with the models presented in this work; those having a low [N/H], together with high [C/H] and [O/H]. HE1012-1540 and Segue1-7 show such a trend and this cannot be explained correctly with our models. HE1012-1540 has the lowest [N/H] of the sample and is highly enriched in Na and Mg (see Fig. \ref{star7}). Our models cannot explain a very low [N/H] together with a high [Na/H] because Na is boosted thanks to rotation but rotation also synthesises primary N. None of our models can provide the right observed trend (see examples in Fig. \ref{star7}).

\end{enumerate}

The range of observed abundances is very large; up to $\sim 3$ dex for CEMP-no stars having roughly the same [Fe/H]  (red rectangles on Fig. \ref{star1} or \ref{star7}). The observed abundance patterns can vary strongly from one CEMP-no star to another. We think that this variety might rule out by itself a single scenario that could account for the entire CEMP-no star class. In summary, although our approach cannot explain the whole considered sample, it might provide a solution for a significant part of it, or at least a clue on some ingredients (late mixing and progressive mixing) able to reproduce the observations. This gives some support to the idea that mixing (either progressive or late and brutal) has played an important role on the early and massive star generations.

\section{Discussion}
\label{sec:8}

\subsection{Low energetic supernova and winds}
\label{faint}

There are three constraints in favour of an ejection of just the outer layers of the source star. 

First, as we saw, if overly deep layers are ejected, the $^{12}$C/$^{13}$C ratio is too high compared to the observed values. 

The second constraint is related to the s-process. At this metallicity ($Z = 10^{-5}$) and in non-rotating models, the content in $^{22}$Ne (neutron source) and $^{56}$Fe (neutron seed) is too low for the s-process to occur significantly.
However, including rotation produces extra $^{22}$Ne, which significantly boosts  the s-process. \citep[the production of Sr can be raised by several orders of magnitude; ][especially their Fig.~8]{frischknecht16}. The weak s-process mainly happens at the end of the core He-burning phase. At the end of evolution, most of the s-elements are located in the He-burning shell. If the He shell (plus eventually deeper layers) is ejected, the ISM will be enriched in s-elements, contradicting the fact that CEMP-no stars
generally show no s-elements.
For instance, we have seen that HE2331-7155 is best explained by the 20s7 model. This model will produce s-elements because of fast rotation. However, HE2331-7155 shows no enhancement in s-elements ([Sr/Fe] $=-0.85$ and [Ba/Fe] $=-0.90$). As a consequence, only the outer layers (above the He-shell) of the 20s7 model have to be ejected, otherwise some s-elements are released in the ISM. 

Finally, in the late mixing region, a lot of protons are ingested by the He-burning shell. It produces extra energy and inflates the star in this region. This decreases the gravitational binding and makes the outer layers easier to eject.

In the present work, the outer layers of the models are expelled through a low energetic supernova. The CEMP-no stars form from that material. Our source star models lose almost no material through winds. It might be, however, that the winds are underestimated. \cite{moriya15} have shown that the envelope of massive Pop. III stars can become pulsationally unstable near the end of their evolution and then undergo extreme mass-loss events. This might be viewed as a fourth piece of evidence in favour of an ejection of only the outer layers of the source star. Source star models including this new wind prescription should be computed in the future.
We can speculate that a wind occurring at the very end of the evolution will barely change the results presented here since at that time the composition of the outer layers does not change anymore.

If the winds were to occur earlier,  the results discussed here would change for the following reasons: Firstly, the chemical composition of these outer layers would be different from those they would have acquired if that matter were to have remained locked in the star until the end of evolution. Secondly, the importance and even the presence of the late mixing process invoked here might be disputed. This point would require some exploratory work with different mass-loss algorithms. This will be done in a future study.

\subsection{Physical origin of the late mixing process and partially processed zone}
\label{late}

We have seen how a late mixing event in the source star could provide a solution for some CEMP-no stars. In our models, this mixing is artificially triggered. Our conclusions would be of course strengthened if a known physical process were able to explain this mixing. Our main aim here was to see to what extent the CEMP-no sample could be reproduced by standard source-star models and if it could not, to try to find which ingredients are missing.

We note that the treatment of the convection can impact the shell interaction and thus the formation or not of the partially CNO-processed zone. In the Geneva code, the boundaries of the convective zones are determined using the Schwarzschild criterion, and the convection is assumed to be adiabatic\footnote{During the main sequence and core He-burning phase, the core is extended using a penetrative overshoot, the length of which is proportional to a fraction of the pressure scale height at the edge of the core. This is applied neither for the more advanced phases of stellar evolution, nor for the intermediate convective shells, and is thus not relevant to the above discussion.}. The boundaries of the convective zones are sharp (step functions). This prescription probably does not capture the whole physics of convection \citep{arnett15}. Indeed, multi-dimension hydrodynamics numerical simulations of convection in deep stellar interior show that the chemical composition of each side of the convective boundary makes a smooth transition and is not a step function. Moreover, the boundaries are not strict barriers for the chemical elements, and part of them can be mixed through the boundary \citep{herwig06, meakin07, arnett11, cristini16, jones16}, in contrast with the way convection is modelled in the present paper.

It might be that improving the way convection is treated in classical 1D codes to follow more closely the behaviour observed in multi-dimensional simulations strengthens the exchanges between the H- and He-burning shells 
\citep[or even leads to shell mergers. We refer to e.g.][for discussions about shell mergers]{rauscher02, tur10, pignatari15}.
This could induce the creation of the partially CNO-processed zone. If so, the late mixing invoked in our work would simply mean that an overly poor description of the convective boundaries is used in present 1D stellar evolution models.

In any case, mixing between the H- and He-burning shells would be naturally favoured in low-metallicity stars compared to higher-metallicity ones because of the increasing compactness when metallicity decreases and because of the lack of CNO elements that leads to a weaker entropy barrier at the bottom of the H-burning shell.

\section{Conclusions}
\label{sec:9}

We have investigated the origin of CEMP-no stars. The material forming a CEMP-no star could come from a previous massive star, referred to as a source star. 

We have computed $20 - 60$ $M_{\odot}$ source stars with no and fast rotation. Through a comparison between observations and models in the $^{12}$C/$^{13}$C versus C/N plane, we have shown that standard source-star models (rotating or not) have difficulty in providing a material with a high C/N together with a low $^{12}$C/$^{13}$C. Many CEMP-no stars present this trend.
Source-star models tend to produce either a low C/N with a low $^{12}$C/$^{13}$C due to the effect of the CN cycle, or a high C/N with a high $^{12}$C/$^{13}$C because of He-burning that destroys $^{13}$C and $^{14}$N.
Increasing the dilution of the ejected material with the ISM increases the $^{12}$C/$^{12}$C ratio. Decreasing the mass cut $-$ hence expelling deeper layers from the source star $-$ also increases the $^{12}$C/$^{13}$C ratio.

To explain the numerous cases of CEMP-no stars showing  $^{12}$C/$^{13}$C ratios near CNO equilibrium and C/N ratio above the CNO equilibrium value, we suggest that a late mixing, occurring just before ejection is needed. This is the main point of this paper. This conclusion remains robust against changes of the mass cut and dilution that would both increase the $^{12}$C/$^{13}$C ratio much above the observed values. This trend, in our view, reflects a mixing process that is not yet properly accounted for in the stellar models.

The late mixing event should occur between the H- and He-burning shells, a few hundred years before the end of source star's life. This mixing brings extra $^{12}$C from the He to the H shell, boosting the CN cycle in the H shell. The short time remaining before the end of evolution allows $^{12}$C/$^{13}$C to reach its equilibrium value but not C/N. A second possibility to obtain such material would be to undergo a strong mixing event, as described in this work, but not necessarily occurring at the very end of evolution; the mixing would be quickly followed by a dredge up of the partially CNO-processed material up to the surface and then heavy mass loss would occur due to the sudden increase of the surface metallicity. This occurs in the rotating $85\,M_\odot$ model at $Z=10^{-8}$ of \cite{hirschi07} \citep[see also][]{maeder15a}. We plan to further investigate this second scenario in the future.

The generally low $^{12}$C/$^{13}$C ratio observed in CEMP-no stars suggests that only the outer layers should have been expelled by the source star to obtain a $^{13}$C-rich material. This could imply a weak supernova explosion at the end of the source star's life, together with a large amount of matter falling back on the central black hole. Strong winds occurring in late stages is also a possibility. 

We have more closely inspected six CEMP-no stars through a comparison between observed and predicted [X/H] ratios. We found that four out of the six stars probably cannot be explained without a late mixing event in the source star, and that two stars probably cannot be explained without a progressive mixing, achieved by fast rotation of the source star in our models. 
This suggests the possibility of two kinds of mixing operating in CEMP-no source stars: A progressive mixing that could be achieved by rotation, and a late mixing, possibly linked to shell mergers. The late mixing invoked in the present work could simply mean that an overly poor description of convective boundaries is used in current 1D stellar evolution codes.
More generally, we have shown that Na-rich CEMP-no stars are difficult to explain without progressive mixing operating in the source star. Also, $20\,M_\odot$ source stars are generally preferred compared to higher-mass source stars ($32$ or $60\,M_\odot$) because of the lower temperature in the H-burning shell that induces a slower pace of the CN-cycle. 
A few CEMP-no stars could not be explained by either of our source-star models.
The diversity of the abundances observed at the surface of these stars might suggest the need for multiple scenarios.

\begin{acknowledgements} 
RH acknowledges support from the World Premier International Research Center Initiative (WPI Initiative), MEXT, Japan and from the ChETEC COST Action (CA16117), supported by COST (European Cooperation in Science and Technology).
The research leading to these results has received funding from the European Research Council under the European Union's Seventh Framework Programme (FP/2007-2013) / ERC Grant Agreement n. 306901.
\end{acknowledgements}

\bibliographystyle{aa} 
\bibliography{biblio.bib}

\end{document}